# Ultrafast ps–TALIF and streak camera diagnostics of atomic hydrogen in a helium microplasma jet

D. Stefas[1], Y. Agha[1], L. Invernizzi[1], J. Santos Sousa[2], S. Prasanna[1], J. Franzke[3], C. Anastassiou[4,5], G. Lombardi[1], K. Gazeli[1]

[1] *Laboratoire des Sciences des Procédés et des Matériaux (LSPM – CNRS), Université Sorbonne Paris Nord, Villetaneuse, UPR 3407, F-93430, France*
[2] *Laboratoire de Physique des Gaz et des Plasmas (LPGP) - CNRS & Université-Paris-Saclay, Orsay, France*
[3] *Leibniz-Institut für Analytische Wissenschaften – ISAS – e.V., Bunsen-Kirchhoff-Str. 11, 44139 Dortmund, Germany*
[4] *PHAETHON Centre of Excellence for Intelligent, Efficient and Sustainable Energy Solutions, Nicosia 2109, Cyprus*
[5] *ENAL Electromagnetics and Novel Applications Lab, Department of Electrical and Computer Engineering, University of Cyprus, Nicosia 2109, Cyprus*

Email: *kristaq.gazeli@lspm.cnrs.fr*

## *Abstract*

Picosecond two-photon absorption laser induced fluorescence (ps–TALIF) is combined with a streak camera (~1 ps highest temporal resolution) to probe the effective lifetime ($\tau_{eff_{H(n=3)}}$) and absolute density ($N_H$) of atomic hydrogen in the effluent of a helium atmospheric-pressure microplasma jet (µAPPJ). This approach allows for improved temporal resolution compared to conventional nanosecond diagnostics, enabling measurements of $\tau_{eff_{H(n=3)}}$ as small as 50 ps, the corresponding $N_H$ reaching down to $10^{14}$ cm$^{-3}$. Atomic densities are calibrated by performing identical ps–TALIF measurements in krypton gas contained in a custom-built low-pressure cuvette. Physical mechanisms that may be involved in the TALIF scheme such as photoionisation and/or stimulated emission are also assessed, ensuring reliable studies. The determined $\tau_{eff_{H(n=3)}}$ and $N_H$ increase with the helium flow rate ($Q_{He}$) in the range $Q_{He}$=0.3–1 slm. Both quantities are maximized near the exit nozzle of the µAPPJ, obtaining values below 400 ps and $6\times10^{14}$ cm$^{-3}$, respectively. As the axial distance increases, $\tau_{eff_{H(n=3)}}$ declines with a rate of ≈65 ps/mm for $Q_{He}$=1 slm which is about 3 times smaller than for $Q_{He}$=0.3 slm. These findings reveal a strong correlation between the experimentally-measured $\tau_{eff_{H(n=3)}}$ and the local air entrainment into the jet as indicated by their comparison with calculated effective lifetimes based on published quenching rates. Furthermore, operating the µAPPJ in the burst mode allows for the estimation of the residence time ($t_{res}$) of ground state H–atoms in the helium gas channel, which is larger for $Q_{He}$=0.3 slm ($t_{res}$≈1.2 ms) compared to $Q_{He}$=1 slm ($t_{res}$≈0.5 ms). H-atoms consumption in the gas channel can be affected by diffusion and mechanisms involving neutral ground-state and metastable species among others. Furthermore, based on an error propagation analysis, density uncertainty as high as 64% (depending on the operating condition) is revealed. This mainly originates in the ratio of the two-photon absorption cross sections of Kr and H atoms ($\sigma^{(2)}{}_{Kr}/\sigma^{(2)}{}_{H}$) which has not yet been measured in the picosecond regime. Nevertheless, the combined utilisation of ps–TALIF and streak camera diagnostics demonstrates high sensitivity and temporal resolution for directly probing key reactive species in atmospheric pressure microplasmas.

## 1. Introduction

Two Photon Absorption Laser Induced Fluorescence (TALIF) is a diagnostic that enables non-invasive and selective measurements of key atomic species (e.g., H–, N–, O–atoms) in plasmas [1–3]. To achieve this, the effective lifetime ($\tau_{eff}$) of the laser-excited atomic states is among the key quantities to be experimentally determined. The use of conventional nanosecond TALIF (ns–TALIF) allows directly accessing atomic lifetimes and densities in plasmas operating in controlled gas mixtures at low-to-moderate pressures (typically up to several tens mbar [4,5]). However, ns–TALIF reaches its limits when applied to plasmas operating in more complex gas mixtures and higher pressures. A characteristic example refers to atmospheric pressure plasma jets, where the nature and density of different quenchers in the effluent are difficult to determine due to a non-uniform mixing of air impurities with the plasma effluent. In these plasmas, the $\tau_{eff}$ of the laser-excited states may fall to sub-ns timescales depending on the species studied, the quenchers' nature and their density [6–10]. Under these conditions, the pulse duration of common ns lasers (typically several ns) exceeds the $\tau_{eff}$ of the excited species generated, impeding the direct extraction of the corresponding actual fluorescence signals. Thus, ns–TALIF essentially becomes impractical for direct measurements of $\tau_{eff}$ and absolute densities of reactive atoms in these plasmas without a previous consideration of relevant models, knowledge of accurate densities of major quenchers and corresponding quenching rates [4,11,12].

Recently, picosecond (ps) TALIF (ps–TALIF) has emerged as a promising extension of ns–TALIF for directly probing reactive atoms (such as H–, N–, and O–atoms) in plasmas [3,6–10,13–15]. This is because the duration of ps lasers is significantly smaller than the decaying phases of common TALIF signals. A factor that affects the reliable use of ps lasers in TALIF studies is the laser intensity (units: W/cm$^2$) at the location of the measurement. This must be kept small enough to avoid saturated fluorescence regimes due to a substantial depletion of the species ground state by the laser, and the laser-excited states by secondary physical processes such as photoionization and stimulated emission [1–3]. A common practice to avoid saturation is to ensure that the TALIF signal's intensity scales quadratically with the laser intensity (the so-called quadratic regime). In the quadratic regime the depletion of the ground state's density by the laser should be negligible, and that of the fluorescing state should be dominated by collisional quenching and spontaneous emission (fluorescence) [2].

Particularly, ps-TALIF seems ideal for application in atmospheric pressure plasmas [6–10,13,16]. For instance, Schröter *et al.* used ps-TALIF (30 ps laser pulse; 10 Hz repetition rate) to determine absolute densities of atomic hydrogen and oxygen at the exit nozzle of the COST radio frequency (RF) jet operated in He/H$_2$O mixtures [6]. Calibration of TALIF signals was done by performing identical measurements in Kr and Xe gases, respectively, contained in a low pressure (≤13.3 mbar) cuvette. For all species, the criterion of the quadratic regime was satisfied, providing adequate laser energies to avoid TALIF saturation. The fluorescence signals were captured with a fast ICCD camera using variable gate widths (2–10 ns) and steps (0.2–1 ns) depending on the species probed. The choice of these gate widths and steps ensured a good signal to noise ratio of the recorded TALIF signals and did not affect the measured $\tau_{eff}$ of the laser-excited states. This study reported detailed effective decay rates and quenching coefficients of the laser-excited states by H$_2$O. Indicative $\tau_{eff}$ values of H– and O–atoms measured in a He/H$_2$O mixture (500 sccm He – 1240 ppm H$_2$O) were as low as 0.8 ns and 7.2 ns, respectively. The addition of H$_2$O in the plasma affected the absolute densities of both species, the highest densities being 4.3×10$^{13}$ cm$^{-3}$ (O–atom) and 6×10$^{14}$ cm$^{-3}$ (H–atom). Klose *et al.* used the same laser system, calibration method and detector as in [6] and performed ps-TALIF measurements of H– and O–atoms in an RF-driven kINPen-sci plasma jet (gas mixture: Ar/H$_2$O) [7]. The $\tau_{eff}$ of the laser-excited O–atom was measured to be as low as 1.73 ns at an axial distance z=0.25 mm from the exit nozzle using a 5 ns gate width and a 0.5 ns step on the ICCD camera. However, the $\tau_{eff}$ of excited H–atom was estimated (90±14 ps) to be lower than the detection limit of the system (about 100 ps) and a calculated value (based on reported quenching rates) was alternatively used to derive its density. For both species, maximum absolute densities were localised at z=0.25 mm: $3.5\times10^{15}$ cm$^{-3}$ (H–atom) and $3.8\times10^{15}$ cm$^{-3}$ (O–atom), showing asymmetric radial decreases. Besides, both densities were no longer measurable beyond z=4 mm, being smaller than the system's detection limit ($\approx 10^{15}$ cm$^{-3}$).

The spatial mapping of ground state oxygen in the effluent of an RF-driven COST jet, operated in He, He/O$_2$, and He/H$_2$O and interacting with a liquid surface, was reported by Myers *et al.* [8]. TALIF was performed with a ps laser (30 ps pulse width; 50 Hz rate) backed with a ns ICCD camera as the fluorescence detector. The density calibration was done by introducing small admixtures of Xe into the He jet (the plasma being switched OFF) and capturing corresponding TALIF signals. The quadratic regime was verified for both species. The $\tau_{eff}$ of laser-excited O–atoms measured in the absence of the liquid were found to be smaller than 1 ns (indicative value at the tip of the effluent in He—1%O$_2$ mixture). These measurements improved by about 30% the accuracy of the densities obtained compared to the case of using extrapolated quenching rates from low pressure studies. The maximum O–atom densities depended on the O$_2$ and H$_2$O contents into the jet and varied between $7.5\times10^{15}$ cm$^{-3}$ (pure He) and $3\times10^{15}$ cm$^{-3}$ (He mixed with 1%O$_2$). When the jet interacted with the liquid target, a significant reduction on the O–atom density was measured near its surface compared to the density measured at the same distance in the target-free jet. In another work, Brisset *et al.* [13] investigated the axial and temporal variation of atomic hydrogen and oxygen in a pulsed-driven pin-to-pin He–H$_2$O discharge using the same ps-TALIF and detector system as in [6]. The effective lifetimes of laser-excited O– and H–atoms were as low as 9 and 1.7 ns. For a voltage amplitude of 2 kV and a gap of 2.2 mm, the densities of O– and H–atoms exhibited noticeable temporal gradients in the afterglow depending on the water content into jet, reaching peak values of $3\times10^{16}$ cm$^{-3}$ (O–atom; mixture: He—0.25%H$_2$O) and $10^{16}$ cm$^{-3}$ (H–atom; mixture: He—0.25%H$_2$O), respectively, both measured 1 µs after the discharge. Those measurements were coupled to a 1D fluid model shedding light to important reaction pathways leading to the generation of O– and H–atoms especially in the discharge afterglow. Finally, Khan *et al.* showed that it is possible to perform density measurements with ps-TALIF in partially-saturated fluorescence regimes [16]. The species of interest was atomic nitrogen generated in a RF-driven plasma jet ignited in pure Ar and different Ar/N$_2$ mixtures. The ps laser had a pulse width of 30 ps and a repetition rate of 50 Hz. The calibration of the N–atom densities was done by performing TALIF in Kr (1% added in Ar gas). The effective lifetime of the laser-excited N–atoms along the axis of the plasma effluent was measured with an ICCD camera (2 ns gate and variable step). The $\tau_{eff}$ varied between 1 ns (Ar/N$_2$) and 1.5 ns (pure Ar) in a good agreement with corresponding theoretical calculations performed using available quenching rates. Maximum densities up to about $4\times10^{14}$ cm$^{-3}$ were measured either close to the exit nozzle or further away from it depending on the axial distance and the gas composition.

The combination of ps-TALIF with conventional detectors exhibiting temporal resolutions in the ns timescale may limit the precision of the diagnostic when applied to atmospheric pressure plasmas [7]. Ideally, directly capturing ultrafast decay times of excited atoms in these plasmas requires detectors with exceptional temporal resolution. A streak camera is a representative ultrafast detector, reaching temporal resolution as low as a few ps or even lower. Therefore, backing ps-TALIF with a streak camera is a perfect choice for recording fast decay rates of fluorescing atoms in reactive plasmas. However, the reliable implementation of streak cameras in ps-TALIF studies requires careful consideration of their peculiarities. A streak camera can capture light events within different time windows (also known as time ranges). Invernizzi *et al.* [9] and Stefas *et al.* [10] used a streak camera to capture raw ps–TALIF signals of H–atoms in two different plasma sources and also characterize the properties of the laser pulses used. The smallest available time range of the streak camera was 100 ps (corresponding to a temporal resolution of ~1 ps) allowing to capture, e.g., a laser pulse duration as low as 6 ps (value measured around 205 nm). However, this time range was not sufficient to fully enclose the raw fluorescence signals generated in the plasmas. This was because the TALIF signals exhibited larger durations and, thus, they were recorded using larger time ranges. In these time ranges, a distortion of the raw TALIF signals was observed due to the apparatus function of the streak camera (more details can be found in [9]). To extract actual TALIF signals and obtain more accurate effective lifetimes of the laser-excited H–atoms, a methodology was developed allowing the removal of the instrumental function of the detector from the raw TALIF signals. This methodology was applied by Siby *et al.* [15], where ps-TALIF and streak camera diagnostics were used to determine the $\tau_{eff}$ and the absolute densities of H–atoms in a microwave plasma torch generated in controlled $H_2$ pressure (20–125 mbar). It was shown that the effective lifetime of H–atoms is affected by the operating pressure reaching values less than 200 ps even at intermediate pressures (125 mbar).

In this work, we employ ps–TALIF and streak camera diagnostics to perform direct measurements of effective lifetimes ($\tau_{eff}$) and absolute densities ($N_H$) of reactive H–atoms generated in a helium atmospheric-pressure microplasma jet (µAPPJ; testbed source). This system allows for measurements of $\tau_{eff}$ under conditions where conventional nanosecond diagnostics (ns–TALIF, ICCD) reach their limits. Atomic hydrogen is an intermediate species participating in key reactions taking place in low–, moderate– and high–pressure plasmas [6,11,13,15,17,18]. However, its direct quantification in moderate– and atmospheric–pressure plasmas is challenging because the $\tau_{eff}$ of laser-excited H–atoms may become smaller than 1 ns [6,7,15] and, in some plasmas, it can even fall below 100 ps, being close to the detection limit of conventional detectors [7,15]. This is where the streak camera employed here becomes beneficial. Its combination with ps–TALIF allows directly determining $\tau_{eff}$ and $N_H$ being as small as 50 ps and $10^{14}$ cm$^{-3}$, respectively, in the µAPPJ effluent. The description of the plasma source, experimental setup, diagnostic tools and methods applied is detailed in section 2. Section 3 presents the results obtained: discharge electrical and optical emission features, TALIF in Kr gas (used for H–atom density calibration), H–atom effective lifetime and absolute densities measured. Discussion of the results obtained is done in section 4 followed by the main conclusions in section 5.

## 2.   *Experimental setup*

### 2.1.   *µAPPJ source, ps-laser system and streak camera*

The experimental setup used to perform ps-TALIF measurements of H–atoms in the µAPPJ is schematically shown in **Fig.1**. The laser system consists of three units (EKSPLA®): *(i)* a picosecond (ps) Nd:YLF pump laser (PL3140, $\lambda_{Laser}$=1053 nm, 10 ps pulse duration, 5 Hz repetition rate, 40 mJ pulse energy) with a master oscillator, a regenerative amplifier, and a power amplifier, *(ii)* a harmonics unit (APL2100) which amplifies the laser pulses and performs fundamental frequency doubling (527 nm) and tripling (351 nm), and *(iii)* an optical parametric generator (PG411) comprising a series of OPO, OPA, SH and DUV crystals in order to generate ps laser pulses in the wavelength range 193—2300 nm (tuning step: $\Delta\lambda_{Laser}\geq$1 pm) with a typical duration between 4 and 6 ps as per EKSPLA®. We also used a streak camera (Hamamatsu C10910-05) to confirm this value. A brief description of its usefulness is provided in this section while more details about its operating principles and peculiarities for implementation in ps-TALIF studies and laser pulse measurements, are given in refs. [9,10].

A streak camera is a detector making it possible to capture ultrafast light events with a temporal resolution <800 fs (e.g., model C16910 from Hamamatsu Photonics). Typically, it consists of an entrance slit followed by focusing optics, and a streak tube comprising a photocathode, a mesh electrode, a sweep circuit, a micro-channel plate (MCP), and a phosphor screen. Once a light event/impulse passes through the slit, it is focused on the photocathode where the incident photons are converted to electrons. These are then accelerated by the mesh electrode towards a sweep unit containing two parallel plates. The plates are biased by a high voltage ramp which fastly sweeps the photoelectrons in a direction from top to bottom. Then, the photoelectrons are deflected at different times and angles as they are being conducted to the MCP where they are multiplied to finally hit the phosphor screen. There, they are converted back into light that is recorded by a readout complementary metal oxide semiconductor (CMOS) camera (ORCA®-Flash4.0 V3 C13440-20CU). The vertical direction of the phosphor screen serves as the temporal axis, while the other axis contains the location of the incident light in the horizontal direction.

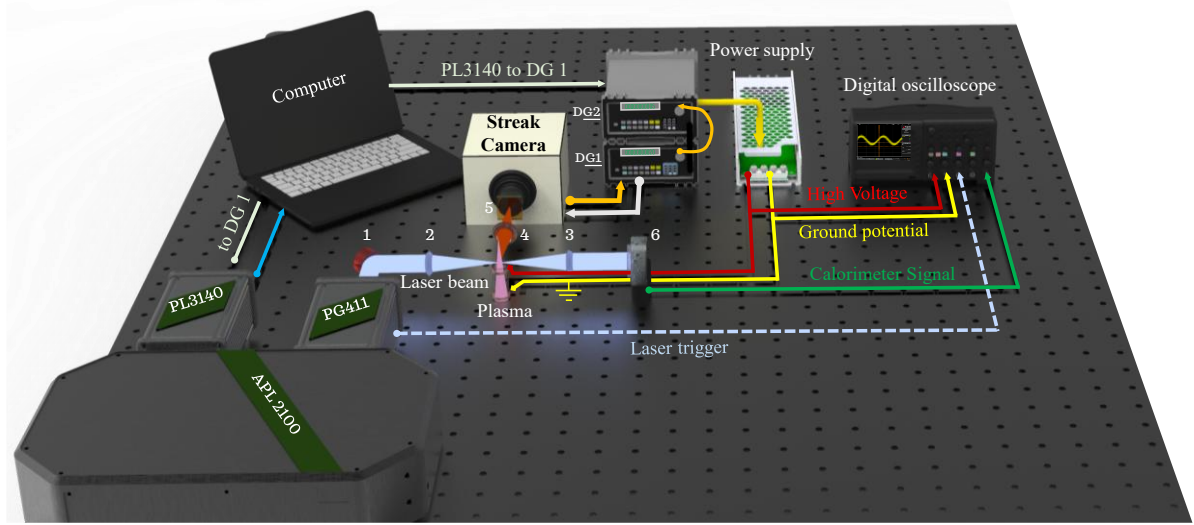

**Fig.1**: *Experimental setup of the laser system, plasma device, TALIF collection optics, streak camera, power supply, delay generators and oscilloscope (1: UV mirror, 2 and 3: f/35 cm UV lenses, 4: f/10 cm lens, 5: bandpass filter, 6: pyroelectric sensor).*

The sweep time (also known as time range, TR) of the camera can be varied between 100 ps and 1 ms in dedicated computer software. The TR chosen is directly connected to the steepness of the voltage ramp applied to the sweep circuit in the streak tube. Thus, it is necessary to select a TR which is sufficiently larger than the total duration of an incoming light event to be captured. The highest temporal resolution of our camera (~1 ps) is reached using a TR=100 ps, which, as already mentioned, is here useful to measure the laser pulse duration. To achieve this, a delay generator (Stanford Research Systems DG645; ≈20 ps jitter) is used to trigger the streak camera which is operated in the single-shot mode. The laser system provides a transistor-transistor-logic (TTL) pre-trigger (PRET) signal to trigger the DG645 with an adjustable delay and a low jitter (2-3 ps). The slit width of the camera, which also affects its temporal resolution [**9**], is set to 10 μm, and a reflection of the laser beam is sent to the slit through a glass diffuser (Thorlabs DG10-600-P01) to avoid damaging the photocathode. For performing the single-shot measurements of the laser pulses, a jitter correction process is directly performed in the camera software.

The energy of the laser pulses is continuously monitored end-on behind the beam's interaction volume with the plasma using a pyroelectric sensor (Coherent J-10MB-LE) connected to a digital oscilloscope (Teledyne Lecroy WaveSurfer 10; 1 GHz, 10 GS/s). Different neutral density filters with variable transmissions (3–92%) can be placed after the output of the PG411 to adequately reduce the laser energy.

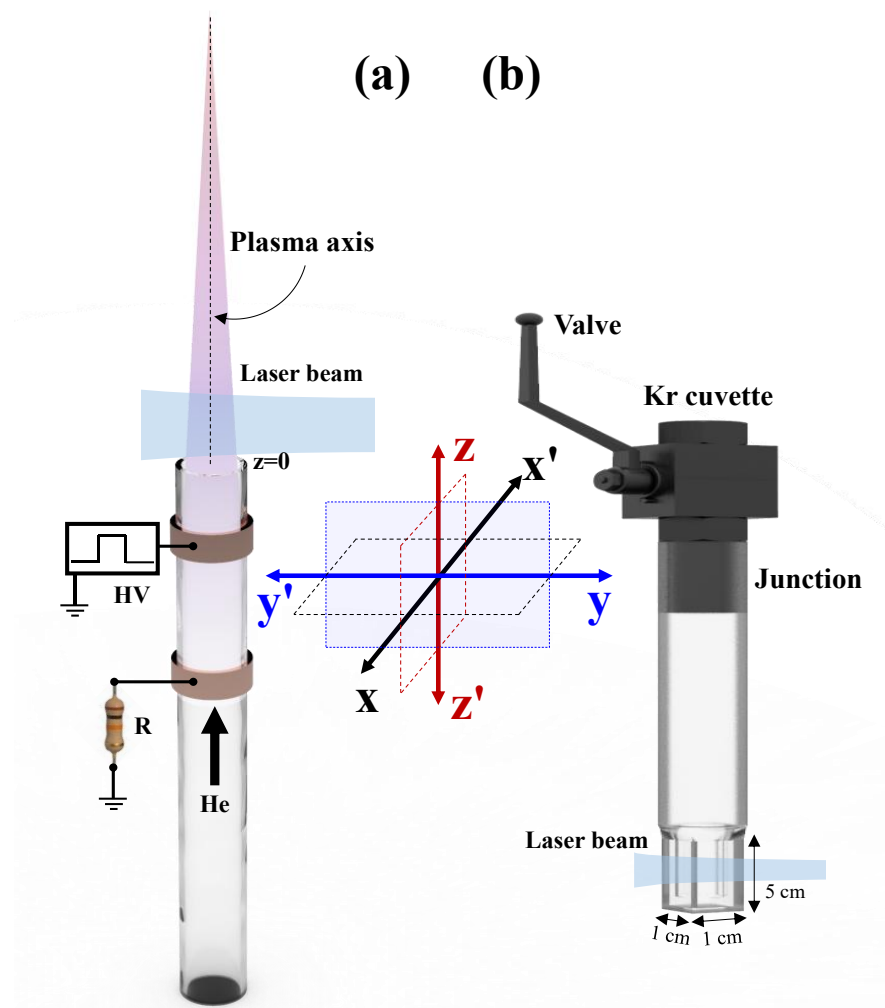

***Fig.2****: (a) µAPPJ and (b) custom-built cuvette used for TALIF calibration (schemes not in scale).*

The plasma device is a µm scale atmospheric pressure plasma jet (µAPPJ; **Fig.2a**) which has been proposed as a soft ionisation source for applications in analytical chemistry [19–24]. It consists of a borosilicate glass microtube (450 µm inner diameter; 900 µm outer diameter; 30 mm length) which acts as a dielectric material to limit the current flowing between the electrodes and guide the operating gas (He) in the surrounding ambient air. Two aluminum ring electrodes are soldered on the outer surface of the microtube having a distance of about 10 mm between them. The upper electrode (1 mm away from the exit nozzle) is the powered one, while the lower electrode is grounded. This setup represents a dielectric barrier discharge (DBD) configuration. The operating gas is pure He from Air Liquide (≥99.999% purity; certified impurities: ≤3 ppm $H_2O$, ≤0.5 ppm $C_nH_m$ and ≤2 ppm $O_2$). Helium is supplied to the device through a polyurethane flexible tube connected to a mass flow regulator (Bronkhorst MV-392-He; $Q_{MAX}$=2 slm; ±2% precision); This allows adjusting its flow rate ($Q_{He}$) between 0.3 and 1 standard liters per minute (slm). In this study, helium flows upwards as shown in **Fig.2a**.

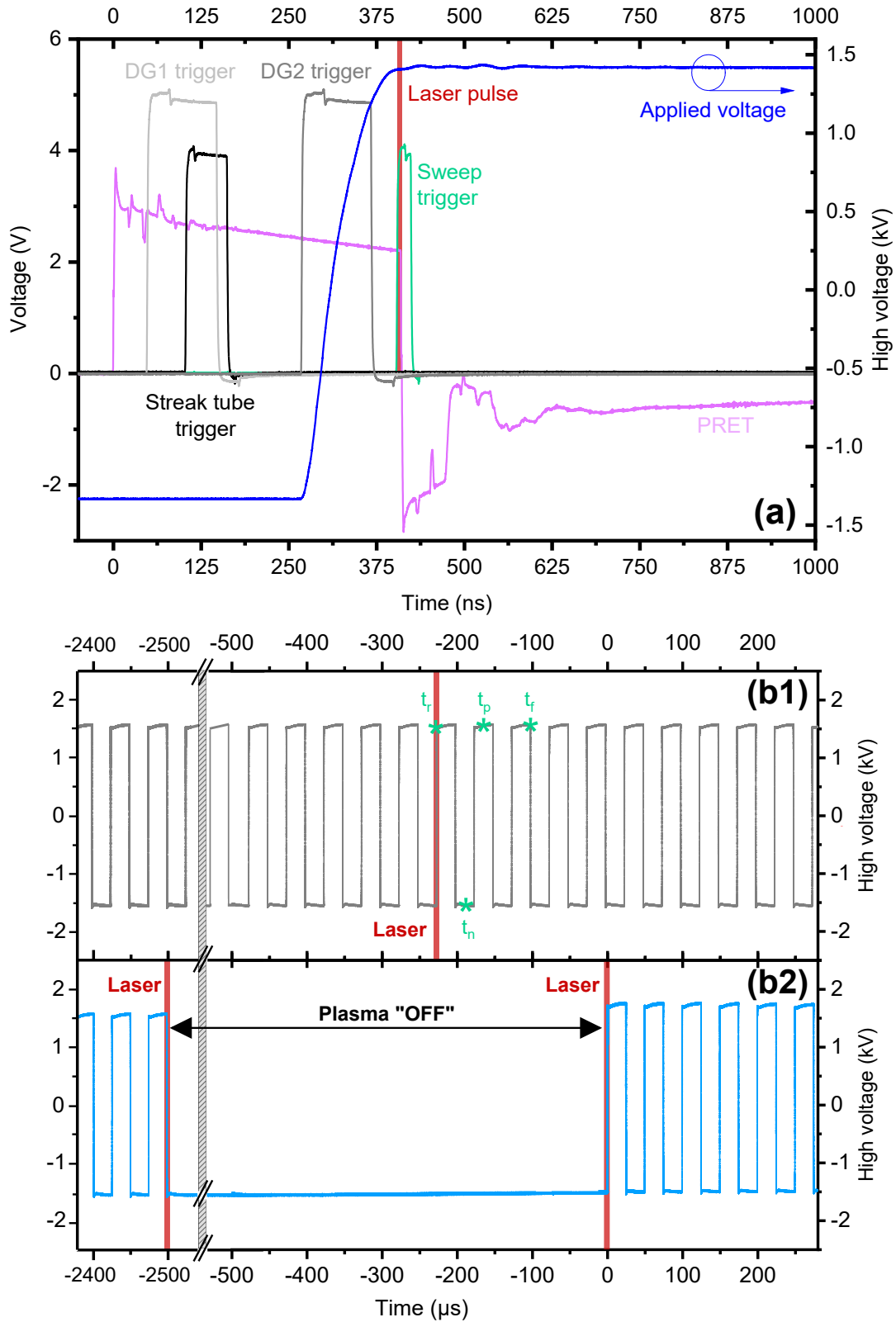

*Fig.3a*: Timeline of signals used to trigger the streak camera and probe H–atoms at distinct instants within consecutive HV pulses. *b1*: typical 20 kHz sequence of HV pulses applied to the µAPPJ; the laser (red) can be selectively shot at any instant, e.g., within the rising ($t_r$) or falling ($t_f$) phases, or the negative ($t_n$) or positive ($t_p$) plateau of the same or different HV pulses. *b2*: burst mode operation comprising 3900 HV pulses per burst at a repetition rate of 20 kHz.

To ignite the plasma, bipolar square high voltage (HV) pulses are applied between the electrodes (**Fig.3**). These are generated by a custom-built power supply and exhibit the following characteristics: 3 kV$_{pp}$ (100 ns rise/fall times), 20 kHz repetition rate, and 24 µs pulse width. These features are kept constants for all the experiments performed here, being monitored in the oscilloscope by means of a passive HV probe (Teledyne Lecroy HVP120; 400 MHz). The total current (I$_{Total}$) in the DBD, i.e., the tube's region extending between the electrodes of the µAPPJ, is monitored in the oscilloscope by measuring the voltage drop across a shunt resistor (**Fig.2a**). This is achieved by using another passive voltage probe (Rohde & Schwarz RT-ZP10; 500 MHz). The recorded HV and current signals are averaged over 100 periods in the oscilloscope.

The µAPPJ is mounted in a 3-dimensional (x, y, z) translational stage with the microtube axis being normal to the optical table. In this way, TALIF measurements of H–atoms are performed at different heights along the propagation axis (z) of the effluent by starting just after the microtube nozzle (z=0 mm; **Fig.2a**). This is done under steady state conditions by applying a constant time delay between the PRET and the instant of the laser pulse generation. The PRET signal is used as the master clock (5 Hz repetition rate), thus driving suitable digital delay/pulse generators (DG; Stanford Research DG645 or Hamamatsu jitter-free delay unit C1097-05) and synchronising the laser, the streak camera, and the pulsed power supply (**Fig.3a**).

To perform time-resolved measurements of H–atoms, the beam's spot position is fixed at z=0 mm (**Fig.2a**). As shown in **Fig.3a**, the PRET signal (magenta) triggers the DG1 which controls the instant of the laser pulse generation by imposing an adjustable delay with respect to the PRET. The DG1 also controls the gate opening of the streak tube (black) which prepares the sweep unit to apply the voltage ramp and deflect the generated photoelectrons. Finally, the DG1 (light grey) triggers the DG2 (dark grey) which has a double role: *(i)* since the laser system operates at 5 Hz, the PRET cannot directly trigger the power supply which requires a 20 kHz TTL signal; thus, this is done using the DG2 which further helps in adjusting the relative delay between the HV pulses and the laser emission (see, e.g., instants t$_r$, t$_n$, t$_p$, t$_f$ in **Fig.3b1**), *(ii)* the DG2 can generate HV bursts in order to study the temporal evolution of the density of H–atoms during the plasma "ON" and "OFF" periods. Here we used 3900 HV pulses per burst at a repetition frequency of 20 kHz, while the burst "OFF" time (i.e., plasma "OFF") is set to 2.5 ms (**Fig.3b2**), which is found to be sufficient for the H–atoms to be consumed in the gaseous channel under our experimental conditions.

To determine the absolute densities of H–atoms in the plasma, the corresponding TALIF signals are calibrated by also performing ps-TALIF measurements in Kr gas (Air liquide, 99.998% purity) contained in a custom-built quartz cuvette (**Fig.2b**; **section 2.3**). While in the case of low pressure plasmas [**5,14,25–27**] the same chamber can be used for plasma generation and calibration purposes, thus ensuring that the optical path along the laser beam and that of the collected fluorescence remain the same, this is not possible in the present work since the plasma operates in an open air environment. A solution in the case of APPJs would be to directly mix Kr with the carrier gas and perform the calibration [**16**]. However, this does not guarantee that the TALIF signal in Kr is not affected by local density fluctuations due to a non-uniform mixing with the surrounding ambient air. This is particularly true for the effluent of the µAPPJ studied in this work which operates in small gas flow rates. Besides, Kr is an expensive gas, and this calibration solution is not economically viable. Therefore, the cuvette provides a convenient solution in our case, as in [**6**]. The cuvette is adequately sealed by performing a metal-glass junction, thus allowing it to preserve Kr–atoms at a fixed pressure (2 mbar) and acquire reproducible TALIF signals for at least two full days of experiments. To achieve this, the cuvette is first pumped down to 10$^{-6}$ mbar using a turbo-molecular pump (Edwards EXT75DX) assisted by a rotary pump (Pfeiffer Duo 6M), then filled up with krypton at a much higher pressure (∼100 mbar) and finally pumped down again to 2 mbar. This pressure is manually adjusted using a diaphragm valve and monitored via a capacitive gauge (Pfeiffer CCR 363; $1.33\times10^{-3}$ – 13.3 mbar; 0.2% accuracy). Another valve is used to seal the cuvette, allowing for its detachment from the pumping line for TALIF studies. The laser beam properties are kept exactly the same between performing ps-TALIF in the microplasma and the cuvette. An attenuation of the laser beam's intensity on the walls of the cuvette (yy' axis in **Fig.2**; 87% transmittance, measured with the same ps laser) is considered for accurate measurements of the laser energy. Besides, an attenuation of the TALIF signal intensity on the side walls of the cuvette (xx' axis in **Fig.2**; 90% transmittance, measured with a Thorlabs PL202 laser at 635 nm) is also considered for correcting the TALIF intensity.

All TALIF signals are recorded with the streak camera using the following settings: 63 MCP gain, 20 exposures, and 10 s per exposure CMOS integration time. This corresponds to a duration of 200 s (i.e., 1000 laser pulses) for capturing one raw TALIF signal. The measured shot-to-shot laser energy fluctuation does not exceed 10% over 1000 laser pulses. For optimal stability, the laser system operates in controlled room conditions (20 ºC and about 50% relative humidity). Additionally, different TR values are used to collect fluorescence signals of H–atoms (TR=5 ns) and Kr–atoms (TR=200 ns). The use of these values ensures that the full raw TALIF signals fit in the CMOS sensor while maintaining an acceptable SNR and good temporal resolution.

For capturing time- and spectrally-integrated images of the plasma effluent, a conventional camera is used. Furthermore, the emissive species generated by the discharge are identified using a compact spectrometer (Avantes Avaspec-ULS4096CL-EVO; 75 mm focal length; grating: 300 l/mm, 300 nm blaze; 200—1100 nm spectral range; 0.7 nm resolution). The plasma emission, spatially-integrated along the effluent's length, is collected using a UV–IR optical fibre (Avantes FC-UVIR200-1; 200 μm core), guided to the entrance slit (10 μm width) of the spectrometer, and recorded by means of an embedded CMOS linear image sensor in a desktop computer. For time-resolved measurements of the discharge's total emission, determination of its propagation length and velocity, an ICCD camera is used (Princeton Instruments PI-MAX 2) equipped with a UV objective lens (Nikkor). The gate width of the camera is set to 10 ns, the step is 5 ns, and the emission accumulates over 60 HV pulses.

*2.2.    Laser beam properties - TALIF spatial resolution*

The laser beam is focused on a position behind the plasma effluent and the centre of the cuvette (**Fig.2**). This allows limiting undesired saturation effects (photoionisation, photodissociation and/or stimulated emission) due to an excessive laser intensity, which obscure the TALIF measurements [**1–3,6**]. Furthermore, it avoids damage to the walls of the Kr cuvette. To determine the spatial resolution of the TALIF measurements, the laser beam radius ($w$) is measured exactly at the intersection of the laser beam with the vertical central axis of the microplasma and the cuvette. Specifically, a knife-edge progressively blocks the laser beam in the direction zz' (**Fig.2**), which is perpendicular to its propagation axis (yy'), and the pyroelectric sensor measures the corresponding laser intensities [**28,29**]. There are various approaches to analyse the data obtained from this experiment and determine the value of $w$. In this work we used two methods.

First, a suitable function (red line) is fitted to the experimental data (black dots) as shown in **Fig.4**. This method considers a Gaussian beam (true here by >90% as per EKSPLA®), the radial intensity of which ($I(r)$) can be expressed as follows:

$$I(r) = I_0 \exp\left(-\frac{2r^2}{w^2}\right) \quad (1)$$

, where $I_0$ is the maximum intensity (in the absence of the knife-edge) and $r$ the radial coordinate. The experimental data are approximated by an integrated Gaussian function [29] or, equivalently, an error function:

$$I(z) = I_{BG} + \frac{I_0}{2} \int_z^\infty \exp\left(\frac{-2(z-z_0)^2}{w^2}\right) dz = I_{BG} + \frac{I_0}{2} erf\left(\frac{\sqrt{2}(z-z_0)}{w}\right) \quad (2)$$

where $I(z)$ is the intensity measured by the pyroelectric sensor at a position $z$ of the knife-edge relative to $z_0$ (i.e., centre of the Gaussian), and $I_{BG}$ is a background signal measured when the laser beam is totally blocked by the knife-edge. For fitting function (2) to the experimental data, a custom-made Python routine is used from the Scipy library [30]. Then, the Gaussian profile (blue line) is obtained by differentiating the fitted error function, and the value of $w$ is defined as the distance $z$ from $z_0$ where the intensity of the Gaussian falls down to $1/e^2$ (=13.5%) of the $I_0$. Applying this method to our data, the value of $w$ is found to be 390±10 μm.

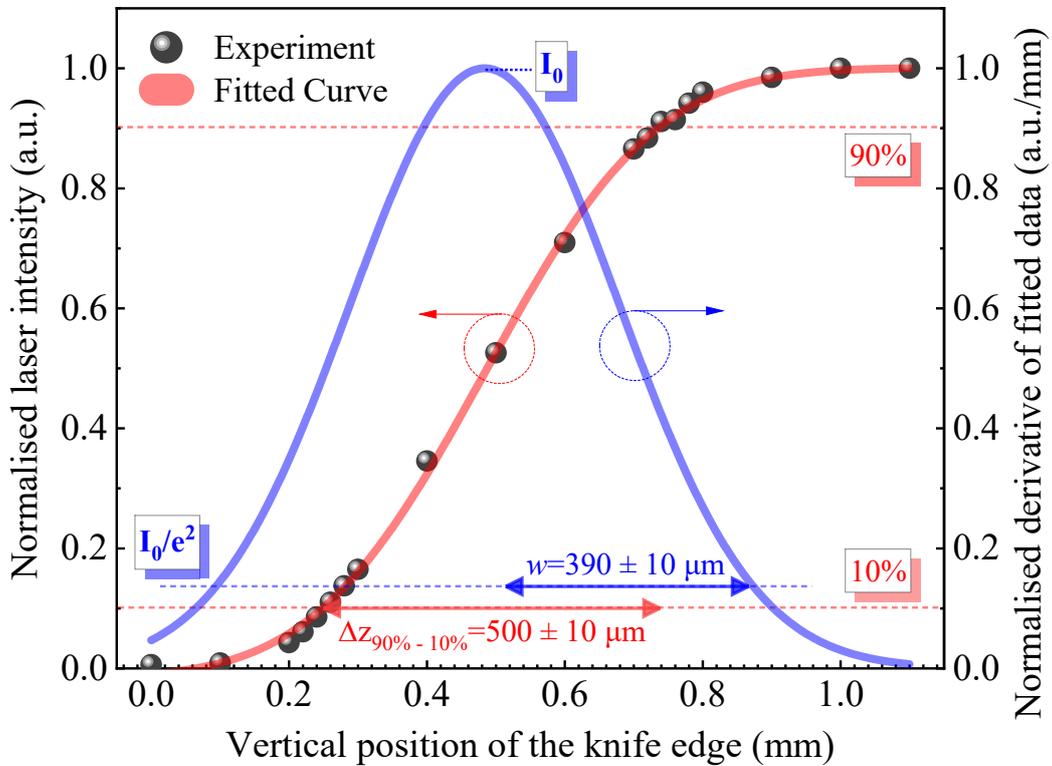

***Fig.4***: *Measured laser intensity (black dots) versus the vertical position z of a knife-edge. The experimental data are fitted with the error function (erf; red). The derivative of the erf is shown in blue (Gaussian profile).*

Second, the difference between the knife-edge positions at which the beam intensity (described by the **eq.1**) falls to 90% and 10% of its peak value ($z_{90}$ and $z_{10}$ respectively) is determined, and the value of $w$ can be calculated as follows [29]:

$$w \approx 0.78\,(z_{90} - z_{10}) \quad (3)$$

The 90%-10% method is in general faster to implement. However, it can be less accurate, particularly when the beam profile significantly deviates from an ideal Gaussian. By using this approach, the value of $w$ is found to be 390±10 μm, agreeing with that obtained by the previous method. Therefore, we may consider that $w≈390$ μm, which corresponds to a full beam width of 780 μm at the centre of the plasma effluent and the cuvette.

## 2.3. H–atoms absolute density measurements

To probe ground-state H–atoms in the plasma effluent, the laser wavelength is set to $\lambda_{Laser}$=205.08 nm (theoretical value) and their excitation to *H(n=3)* is happening with the absorption of two identical UV laser photos (**Fig.5**). The emitted fluorescence signal due to the transition *H(n=3)→H(n=2)* is captured by the streak camera. The time-integrated fluorescence in this case can be expressed with the following equation:

$$S_{F_H} = \eta T V \frac{\triangle\Omega}{4\pi} \frac{A_{3\to 2}}{A_3 + Q_3} \frac{\sigma_H^{(2)} g(v_{res}) G^{(2)}}{(v_{res})^2} N_H \int_0^\infty I_{Laser}^2(t) dt \quad (4)$$

In **eq.4**, $S_{F_H} \propto E_{Laser}^2$ since the laser intensity (W/cm²) is defined as $I_{Laser} = (E_{Laser}/S_{Beam} t_{Laser})$ where $E_{Laser}$ is the laser energy, $S_{Beam}$ the laser's beam area at the TALIF measurement position and $t_{Laser}$ the laser's pulse duration. Furthermore, $\eta$ is the detector's quantum efficiency at the fluorescence wavelength, *T* the transmission of the various optics used (lenses, filters, etc.), *V* the interaction volume of the laser with the probed medium, ΔΩ the solid angle of collection of the fluorescence, $A_{3\to 2}$ the Einstein coefficient describing the fluorescence transition *H(n=3)→H(n=2)*, $A_3$ is the sum of Einstein coefficients for all radiative channels from the laser-excited level *H(n=3)*, $Q_3$ is the quenching rate of *H(n=3)*, $\sigma_H^{(2)}$ the two-photon absorption cross section for the transition *H(n=1)→→H(n=3)*, $g(v_{res})$ the peak value of the two-photon absorption line profile at the resonant laser frequency ($v_{res}$), $G^{(2)}$ the photon statistical factor, and $N_H$ the H–atoms density. The determination of $N_H$ from **eq.4** is challenging particularly due to the difficulties in directly measuring the laser-plasma interaction volume, the solid angle, the two-photon absorption cross-section, and the photon statistical factor [3]. Therefore, a calibration procedure has been proposed which uses identical TALIF measurements in noble gases of well-known densities and allows overcoming these issues [31]. This method has been widely applied for the calibration of atomic densities in plasmas using TALIF [5–7,11,17,18,27].

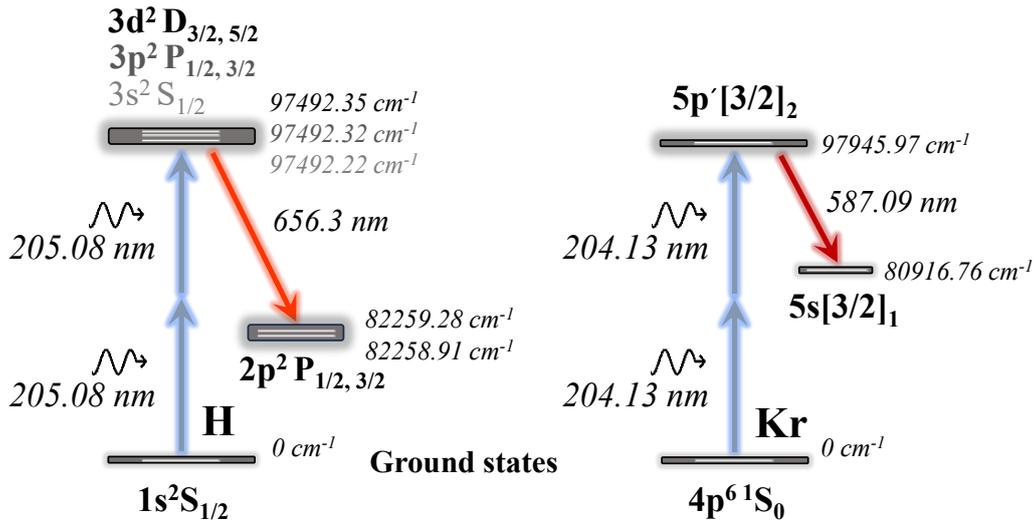

***Fig.5***: *TALIF schemes for H–atoms (left) and Kr–atoms (right).*

In this work, to calibrate the TALIF measurements of H–atoms, TALIF is identically performed in Kr–atoms (theoretical $\lambda_{Laser}$=204.13 nm) contained in the cuvette (**Fig.2b**; 2 mbar). Kr–atom is chosen here as a reference calibration system since the two-photon excitation and fluorescence wavelengths are relatively close to those of H–atom. Furthermore, the optical components used for focusing the laser beam and collecting the fluorescence signals are kept exactly the same for these two atoms. This allows eliminating the experimental complexities related to the direct measurement of $V$, $\Delta\Omega$, $\sigma^{(2)}$, and $G^{(2)}$ since they are essentially the same between the two atoms and can be cancelled out (see **eq.5**). The corresponding TALIF excitation schemes considered for both atoms are shown in **Fig.5**. Following the absorption of two photons by these atoms, the corresponding emitted fluorescence signals from the upper levels are collected perpendicularly (xx' axis; **Fig.2**) to the direction of the laser's beam (yy' axis; **Fig.2**) using an achromatic lens (f/10 cm). Bandpass filters (15 nm bandwidth) with central wavelengths at 656 nm (H–atoms TALIF) and 586 nm (Kr–atoms TALIF) are alternatively placed in front of the entrance slit of the streak camera to filter out any stray light. The interaction volume between the laser beam and the probed medium (plasma effluent or Kr gas) is imaged into the slit of the streak camera. The slit width is set to 100 μm to optimize as much as possible the SNR which is here defined as $SNR = (\underline{S} - \underline{B})/\sigma_{Noise}$, $\underline{S}$ and $\underline{B}$ being the TALIF and background emission intensities, respectively, and $\sigma_{Noise}$ is the standard deviation of the recorded noise.

By rearranging the quantities of **eq.4**, the density of H–atoms ($N_H$) in the ground state can be expressed as a function of the fluorescence signal $S_{F_H}$. This is also done for ground state Kr–atoms in the cuvette, the density of which ($N_{Kr}$) is known. Therefore, by performing the calibration procedure and taking the ratio of densities of the two atomic systems based on **eq.4**, $N_H$ can be expressed as a function of $N_{Kr}$ as follows [**1–4**]:

$$N_H = N_{Kr} \frac{S_{F_H}}{S_{F_{Kr}}} \frac{\tau_{eff_{Kr}}}{\tau_{eff_H}} \frac{g(v_{res})_{Kr}}{g(v_{res})_H} \frac{A_{Kr}}{A_H} \frac{\eta_{Kr}}{\eta_H} \frac{T_{Kr}}{T_H} \frac{\sigma^{(2)}_{Kr}}{\sigma^{(2)}_H} \frac{E^2_{Kr}}{E^2_H} \frac{\lambda^2_{Kr}}{\lambda^2_H} \quad (5)$$

where, for a species $X$ in the ground state ($X$: H–atoms or Kr–atoms), $S_{F_X}$ is the time-integrated TALIF signal acquired by setting the laser's wavelength to the resonant value (i.e., 205.08 and 204.13 nm for H–atoms and Kr–atoms, respectively), $\tau_{eff_X} = 1/(A_{X*} + Q_{X*})$ is the experimentally-measured effective lifetime (or decay time) of the laser-excited states ($X^*$), $g(v_{res})_X$ is the peak value of the two-photon absorption line profile (**section 3**), $A_X$ is the Einstein coefficient of the fluorescence transition of $X$, $\eta_X$ is the quantum efficiency of the streak camera at the fluorescence wavelength (656.3 nm for H–atoms and 587.09 nm for Kr–atoms), $T_X$ is the transmission of the optics at the fluorescence wavelength (including that of the walls of the Kr cuvette), $\sigma^{(2)}_X$ is the two-photon absorption cross-section, $E_X$ is the selected laser energy (**section 3**) and $\lambda_X$ the resonant laser wavelength (**Fig.5**). For both atoms, the absorption line profiles can be approximated with Gaussian functions since they are dominated by the laser linewidth (see **section 3**). In this case, the term $\frac{g(v_{res})_{Kr}}{g(v_{res})_H}$ in **eq.5** equals to the inverse ratio of the corresponding FWHM of the Gaussians, i.e., $\frac{g(v_{res})_{Kr}}{g(v_{res})_H} = \frac{FWHM_H}{FWHM_{Kr}}$ since for a Gaussian: $g(v_{res}) \simeq \frac{0.94}{FWHM}$ [**2**]. The validity of **eq.5** requires that the solid angle of TALIF collection is the same between the two atomic systems. Here, attention is paid to maintaining the same spatial properties of the laser beam and the TALIF collection optics for performing identical measurements in both atoms. Furthermore, TALIF signals for both atoms are collected within the same region of interest of the streak camera sensor (**section 3**).

In **Fig.5** three different excited sub-states of the *H(n=3)* are shown which have energy differences between them $\Delta E < 0.2\ cm^{-1}$. The linewidth of our laser is 6.8 cm$^{-1}$ (29 pm) at $\lambda_{Laser}$=205 nm as per EKSPLA®. Therefore, it could be possible, in principle, to simultaneously excite the three sub-states of *H(n=3)* with our laser. However, based on the selection rules for two-photon laser excitation ($\lambda_{Laser}$=205.08 nm) of ground-state H–atoms to the *H(n=3)* level, only the 3s² S$_{1/2}$ and 3d² D$_{3/2,5/2}$ excited sub-states are populated, the ratio of the two-photon absorption cross sections being $\sigma^{(2)}_{1s\rightarrow 3d}/\sigma^{(2)}_{1s\rightarrow 3s}=7.56$ [32,33]. This means that following the two-photon excitation of ground-state H–atoms, only 11.7% of them will be excited in the 3s² S$_{1/2}$ sub-state, the remaining 88.3% being in the 3d² D$_{3/2,5/2}$ sub-state. The mean natural lifetimes of the 3s² S$_{1/2}$ and 3d² D$_{3/2,5/2}$ sub-states following the radiative transitions 3s² S$_{1/2}$→2p² P$_{1/2,3/2}$ and 3d² D$_{3/2,5/2}$→2p² P$_{1/2,3/2}$ are 159 ns and 15.6 ns, respectively [33]. Taking into account the weighted populations of the 3s and 3d sub-states, the weighted natural radiative lifetime of *H(n=3)* is found to be $\tau_{nat_H}$=17.4 ns in good agreement with refs. [33,34], and it is used to calculate $A_H = 1/\tau_{nat_H}$ in **eq.5**. Furthermore, for laser-excited Kr–atoms, $A_{Kr}$ is calculated through $A_{Kr} = 0.04 * 1/\tau_{nat_{Kr}}$ where the term 0.04 accounts for the radiative branching ratio of the excited level 5p' of Kr–atoms and $\tau_{nat_{Kr}}$ is the natural lifetime of the 5p' state measured in this work (**section 3**).

***Tab.1****: Parameter values of **eq.5** used for H–atoms absolute density determination.*

| | Parameter's value (associated error) [Note below Tab.1] | | | | | | | | |
|---|---|---|---|---|---|---|---|---|---|
| **Atom** | $S_{F_X}$ (a.u) | $\tau_{eff_X}$ (ns) | FWHM$_X$ (pm) | $A_X$ (s$^{-1}$) | $\eta_X$ (%) | $T_X$ (%) | $\frac{\sigma^{(2)}_{Kr}}{\sigma^{(2)}_H}$ | $E_X$ (µJ) | $\lambda_X$ (nm) |
| **Kr** | 37258 (10%)[1] | 20.2 (5%)[3] | 27.5 (10%)[5] | 12.54*10⁵ (1%)[6] | 0.08 (10%)[7] | 0.89 (6%)[7] | 0.62 (50%)[8] | 0.63 (10%)[9] | 204.110[10] |
| **H** | Measured[2] | Measured[4] | 29.5 (10%)[5] | 5.75*10⁷ (1%)[6] | 0.06 (10%)[7] | 0.99 (1%)[7] | | 41.2 (10%)[9] | 205.062[10] |

[1] Fixed value, here experimentally-measured for P$_{Kr}$=2 mbar at $E_{Kr}$=0.63 µJ in the quadratic regime (**Fig.10**).
[2] Different values used depending on the plasma experiments performed (fixed $E_H$=41.2 µJ in the quadratic regime (e.g, **Fig.10**)).
[3] Fixed value, here experimentally-measured for P$_{Kr}$=2 mbar (**Fig.13**) and $E_{Kr}$=0.63 µJ in the quadratic regime (**Fig.10**).
[4] Variable value experimentally measured in this work (**Fig.14a**).
[5] Fixed values here, experimentally-measured for Kr–atoms (P$_{Kr}$=2 mbar, $E_{Kr}$=0.63 µJ) and H–atoms ($E_H$=41.2 µJ) (**Fig.12**).
[6] Ref. [**35**].
[7] Equipment's documentation/datasheets.
[8] Ref. [**34**].
[9] Fixed values here referring to the selected E$_X$ in the quadratic regime for each atom (**Fig.10**).
[10] Experimental resonant wavelengths set in our laser system. Using these values in **eq.5**, a ratio $\lambda_{Kr}/\lambda_H$ =0.995 is obtained which is the same with that obtained when considering the theoretical resonant wavelengths, i.e., $\lambda_{Kr}$=204.13 nm and $\lambda_H$=205.08 nm.

**Tab.1** shows the values of the quantities used in **eq.5** to determine the H–atoms absolute densities. Most of them are directly measured in this work, some are available in the equipment's documentation, and the rest of them are taken from the relevant literature. Particularly, a precise determination of the ratio of two-photon absorption cross-sections ($\frac{\sigma^{(2)}_{Kr}}{\sigma^{(2)}_H}$) and the effective lifetime of laser-excited H–atoms ($\tau_{eff_{H(n=3)}}$) is key for performing accurate TALIF measurements. Here, the ratio $\frac{\sigma^{(2)}_{Kr}}{\sigma^{(2)}_H}$ cannot be directly determined and is taken from ref. [**34**]. However, this value is measured with a ns laser and presents a large uncertainty (50%). Since no improved measurement is available yet, it is still adopted for ns-TALIF and also ps- and fs-TALIF [3,5,6,36]. Furthermore, at atmospheric pressure, the laser-excited level *H(n=3)* is depleted by collisional quenching with ambient air molecules (N$_2$, O$_2$, H$_2$O) [33,34,36–39]. By definition, $\tau_{eff_{H(n=3)}} = 1/(A_{H(n=3)} + Q_{H(n=3)})$, where $A_{H(n=3)} = \sum_i A_{3i}$ is the sum of Einstein coefficients ($A_{3i}$) for all radiative channels from *H(n=3)* to a lower energy level *i*, and $Q_{H(n=3)} = \sum_q k_q N_q$ (*q*=N$_2$, O$_2$, H$_2$O, He) is the pressure- and species-dependent quenching rate of *H(n=3)*. Consequently, depending on the operating pressure and the nature of quenchers, $\tau_{eff_{H(n=3)}}$ may drop down to even sub-ns timescales, which may impede its precise measurement with conventional detectors (e.g., photomultiplier tubes, ICCD cameras) [**7**]. Therefore, the implementation of a streak camera is beneficial for more accurate measurements of $\tau_{eff_{H(n=3)}}$ in the plasma effluent.

## 3. Results

### 3.1. Electrical and optical emission features

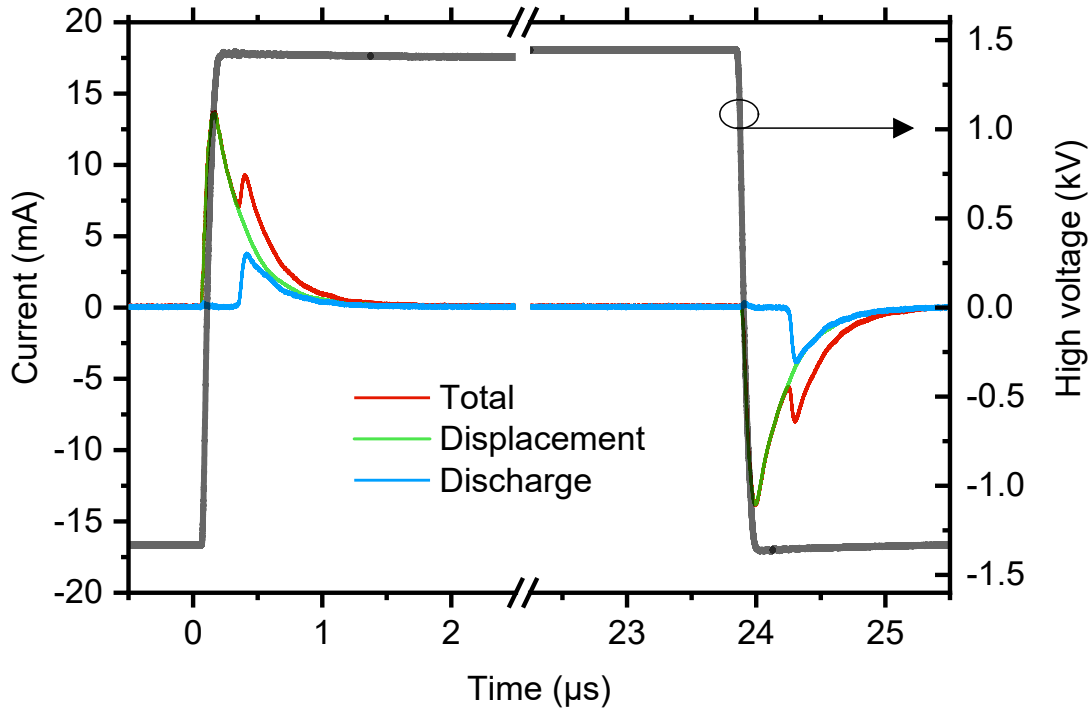

***Fig.6***: *Typical traces of the applied HV pulse and current recorded in the DBD region ($Q_{He}$=1 slm).*

**Fig.6** presents indicative signals of the applied high voltage (grey) and the total current ($I_{Total}$; red) measured in the DBD region. The current consists of the capacitive/displacement current ($I_C$) due to the alternation in the polarisation of the dielectric tube (occurring when the polarity of the applied voltage changes), and the discharge/conduction current ($I_D$) flowing between the electrodes due to the charged species generated when a discharge is ignited. The $I_C$ is assumed to be equal to the current measured when applying voltage to the electrodes and setting $Q_{He}$=0 slm, i.e. helium does not flow through the microtube and, thus, the discharge is switched "OFF". Then, the discharge current ($I_D$; blue) is obtained through a subtraction of the capacitive current ($I_C$; green) from the $I_{Total}$.

In **Fig.6**, two current pulses are generated corresponding to the rising and falling parts of the applied voltage pulse. The former is positive, and the latter is negative. During the positive half cycle, the discharge evolves in three stages following the current evolution, as previously revealed via ICCD imaging [**21–24**]: *(i)* early stage with the ignition of two opposite ionization/excitation waves; the first moves upwards in the helium jet protruding out of the tube (average velocity≈30 km/s), and the second propagates downwards towards the grounded electrode inside the tube (average velocity≈50 km/s), *(ii)* coincident stage starting at the instant of the arrival of the inner wave at the grounded electrode (where $I_D$ reaches its peak value), after which a reflected wave is initiated propagating upwards in the tube towards the powered electrode (average velocity≈50 km/s), *(iii)* afterglow stage, starting once the reflected inner wave reaches the powered electrode. In this work, the use of the ICCD camera allowed us to identify the above-mentioned discharge stages. However, here we are only interested in the length and velocity of the discharge outside the tube since this is the space where TALIF experiments are performed.

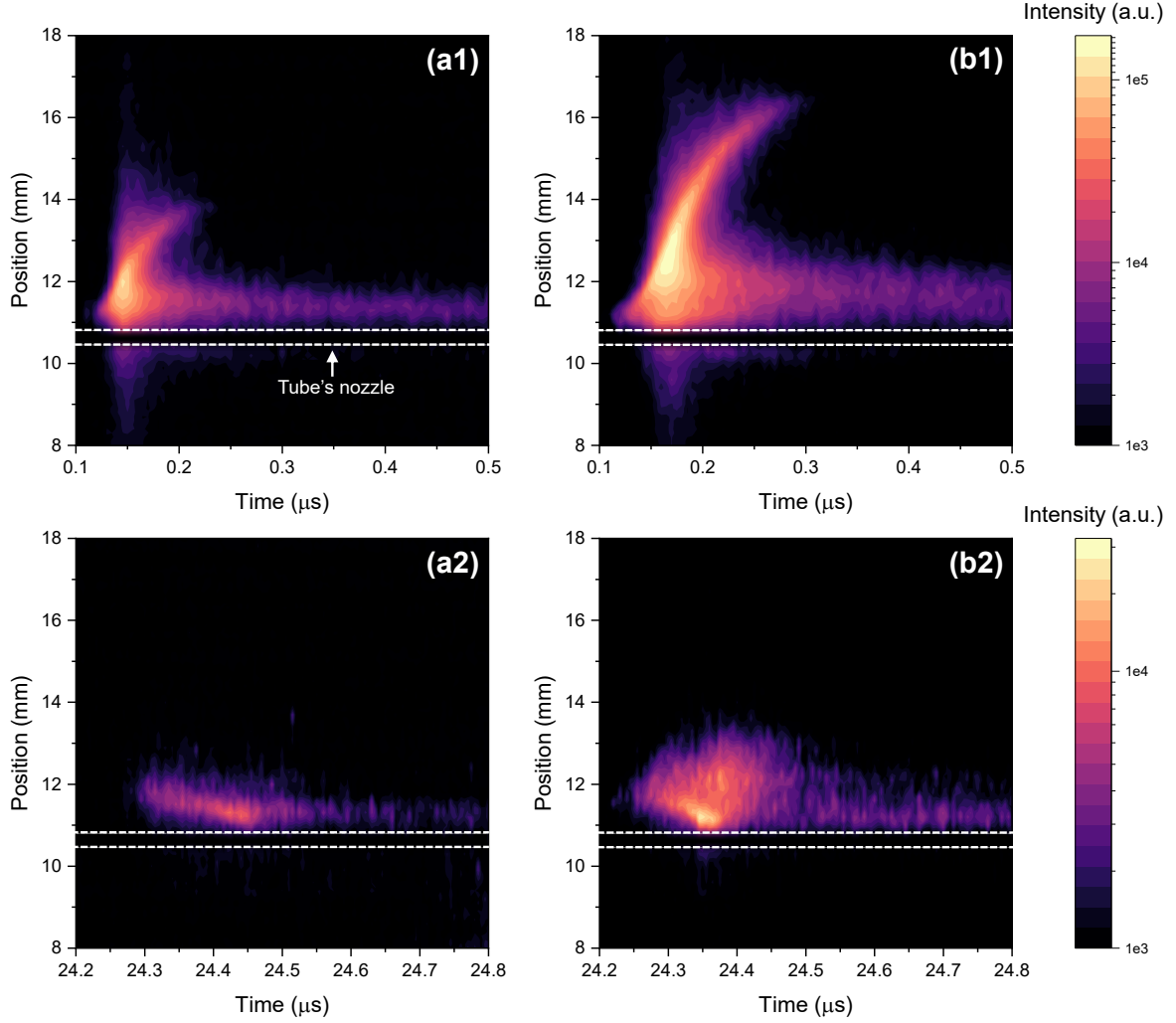

***Fig.7**: Space- and time-resolved images of the discharge's development (total emission) during the early stages at $Q_{He}$=0.3 slm (a1/a2) and $Q_{He}$=1 slm (b1/b2). Images a1/b1 and a2/b2 refer to the discharge events generated during the rising and falling phases of the voltage pulse, respectively.*

**Fig.7** shows the space- and time-resolved total emission profiles of the plasma jet outside the tube at two different He gas flow rates. Besides, a portion of the inner side of the tube is also seen with emission from the downward propagating ionisation wave in the early stage (**Fig.7a1** and **Fig.7b1**). For the discharge outside the tube, during the rising voltage phase, it is characterised by a fast ionisation wave propagation within the first 0.2 μs (**Fig.7a1**; $Q_{He}$=0.3 slm) and 0.3 μs (**Fig.7b1**; $Q_{He}$=1 slm), reaching corresponding lengths of about 3 mm and 5 mm, respectively. After it, a slowly fading stationary emission near the tube's nozzle is recorded. During the falling phase of the voltage pulse the corresponding emission profiles appear much more localised near the tube's nozzle.

It has been shown that the discharge exhibits a much stronger emission intensity in the early and coincident stages, while the afterglow stage is characterised by a long relaxation of excited species generated during the earlier stages. These are emissive atoms and/or molecules (**Fig.8**), as well as non-radiative particles such as metastables (e.g., He ($2^1$S; $2^3$S)), which act as energy reservoirs, affecting the reaction kinetics in the afterglow [**20**]. For instance, He metastables assist the production of charged species (e.g., nitrogen positive ions detected in **Fig.8**) through the Penning ionisation mechanism and may be involved in dissociative reactions as well [**40**].

During the positive half cycle of the voltage (**Fig.7a1** and **Fig.7b1**), charges are accumulated on the inner dielectric walls which are then released and compensated in the falling voltage phase (**Fig.7a2** and **Fig.7b2**) leading to the generation of a negative pulse of the discharge current. In this case, similar stages as before are observed, however, the discharge mechanism refers to a negative streamer propagation versus a positive one in the rising part of the HV pulse. Due to this fact, the outer ionisation wave shows a smaller propagation length, the discharge has a weaker emission intensity, and the velocities of the ionisation waves in the early and coincident stages are smaller.

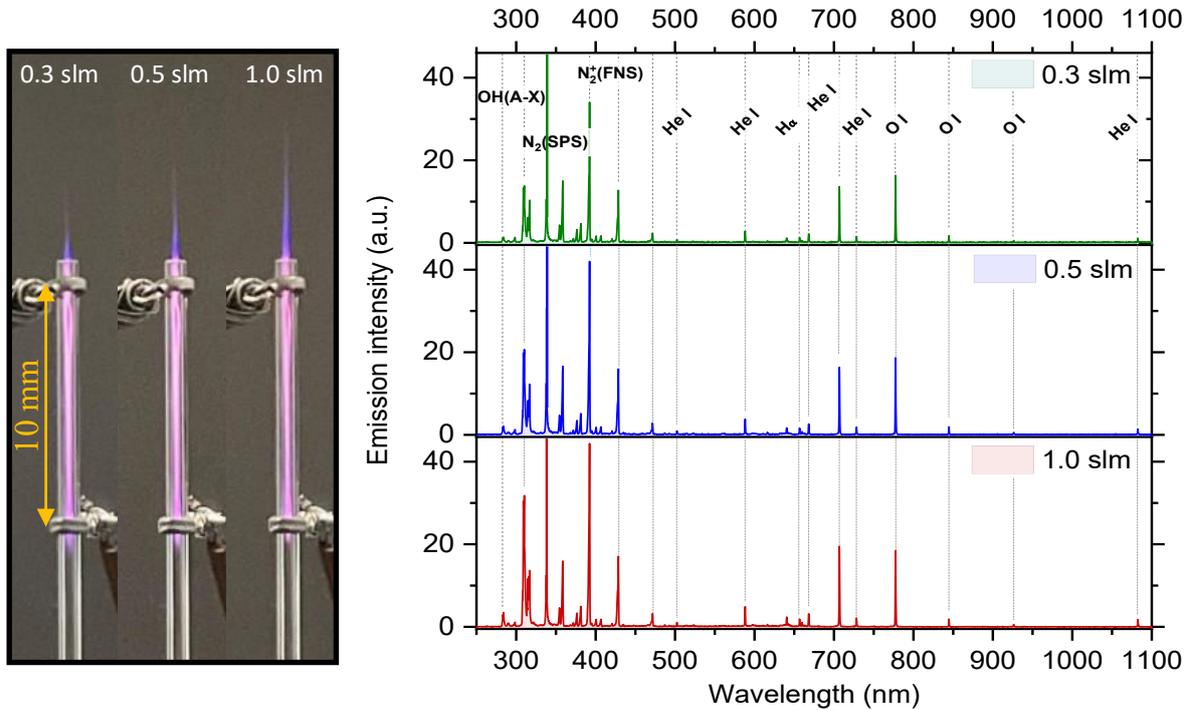

***Fig.8***: *Left: time-integrated photos (f/1.8, 20 ms, ISO-80) of the µAPPJ for different gas flow rates. Right: space- and time-integrated emission spectra (5 s integration time; 10 accumulations).*

**Fig.8** shows time– and spectrally–integrated photos (left) along with optical emission spectra (right) of the discharge for three gas flow rates studied. In both cases the emission properties are dependent on the gas flow rate. Indeed, by increasing $Q_{He}$ from 0.3 to 1 slm, the length of the discharge outside the tube is almost doubled and its light intensity becomes stronger as in **Fig.7**. These observations are further corroborated by the results on the spectrally-resolved plasma emission showing that most intensities of radiative species generated increase with the helium flow rate. For instance, the intensity of OH(A-X) around 306 nm rises from ≈14 a.u. ($Q_{He}$=0.3 slm) to ≈32 a.u. ($Q_{He}$=1 slm), the He I line at 706.5 nm exhibits 1.5 times higher intensity at $Q_{He}$=1 slm, and the Hα emission at 656.28 nm becomes 2.5 times larger at $Q_{He}$=1 slm. Overall, the microplasma source shows a rich chemical reactivity also leading to the generation of other species such as *(i)* excited neutral ($N_2(C)$, $N_2(B)$) and ionic ($N_2^+(B)$) nitrogen molecules emitting between 250 and 900 nm, *(ii)* He I lines between 370 and 1100 nm, some of them producing He metastable states (i.e., He($2^1S$) and He($2^3S$), e.g., through the He I transitions at 501.5 nm and 1083 nm, respectively), *(iii)* atomic hydrogen Hβ emission at 486.13 nm, and *(iv)* O I lines at 777.53 nm and 844.63 nm. The presence of OH, H–atoms and O–atoms in the emission spectrum is evidence of plasma-enabled dissociation of water vapour and $O_2$ molecules which are coming either from the gas bottle (impurities) or the ambient air interacting with the plasma effluent [40]. This is a useful result for the feasibility of capturing TALIF signals of H–atoms in this work.

## 3.2. TALIF in H–atoms and Kr–atoms

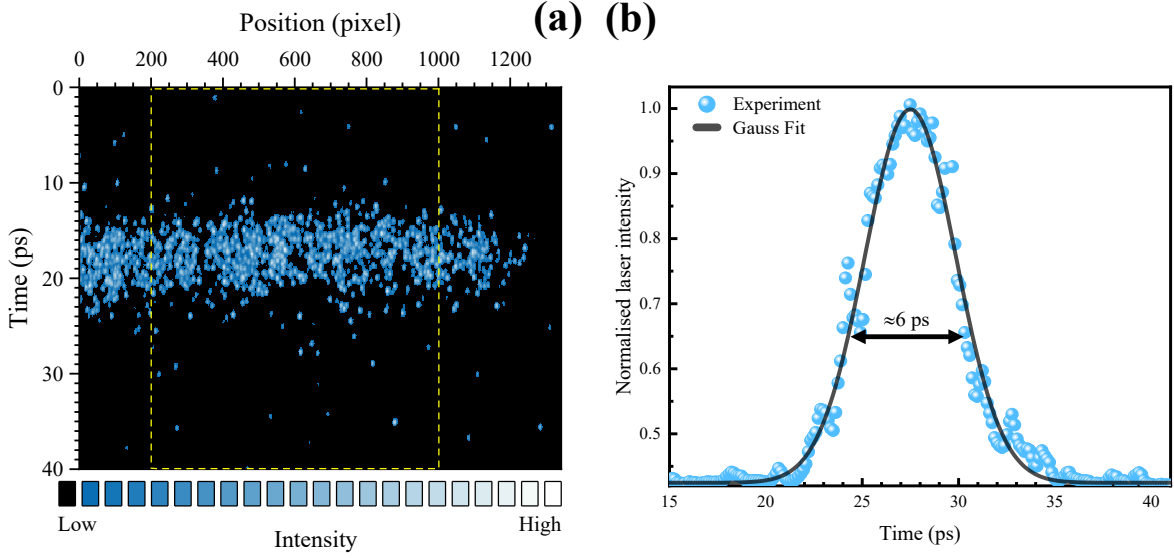

**Fig.9**: *(a) streak image of the laser pulse profile ($\lambda_{Laser}$=205.08 nm; single-shot; TR=100 ps; ~1 ps time resolution; 10 μm slit width). (b) reconstructed laser pulse temporal profile from **Fig.9a** fitted with a Gaussian function (FWHM=(5.8±0.2) ps). An experimental profile is obtained by averaging the spatial intensity between the 200$^{th}$ and the 1000$^{th}$ px of the sensor (yellow box in **Fig.9a**).*

A factor that determines the temporal resolution of TALIF experiments is the duration of the laser pulse ($t_{Laser}$). Here, $t_{Laser}$ was directly measured using the streak camera and a representative measurement of the ps laser pulse is shown in **Fig.9a** ($\lambda_{Laser}$=205.08 nm). The recorded profile in the CMOS sensor is essentially the same between $\lambda_{Laser}$=205.08 nm and $\lambda_{Laser}$=204.13 nm [10]. To maximise the signal-to-noise ratio (SNR), the emission is averaged along the spatial axis of the sensor using a large region of interest (ROI), while we used 20 single-shot jitter-corrected images to build the final streak image (**Fig.9a**). The resulting experimental temporal profile is shown in **Fig.9b**. This is well approximated with a Gaussian function and its full width at half maximum (FWHM) is taken as the pulse duration ($t_{Laser} \approx 6$ ps). The laser's pulse duration, along with its energy ($E_{Laser}$) and the beam's diameter ($S_{Beam}$) at the location of the measurement determine the intensity of the laser ($I_{Laser}$) (**section 2.3**). The $I_{Laser}$ needs to be maintained sufficiently low during TALIF experiments to suppress the influence of secondary physical processes such as photoionization (PIN) and stimulated emission (SE) on the fluorescence signal's intensity [1–4]. As a matter of fact, at decently small laser intensities and pulse durations the depletion of the atomic ground state should be negligible (compared to its total density). This translates to $W_{ge} \times t_{Laser} \ll 1$ where $W_{ge}$ (indices *g* and *e* denote ground and laser-excited states, respectively) is the species' two-photon excitation rate [2]. Furthermore, the depletion of the laser-excited state should be dominated by the fluorescence and quenching processes over PIN and SE. To avoid these issues the fluorescence signal intensity should scale linearly with the squared laser energy which is studied in **Fig.10**.

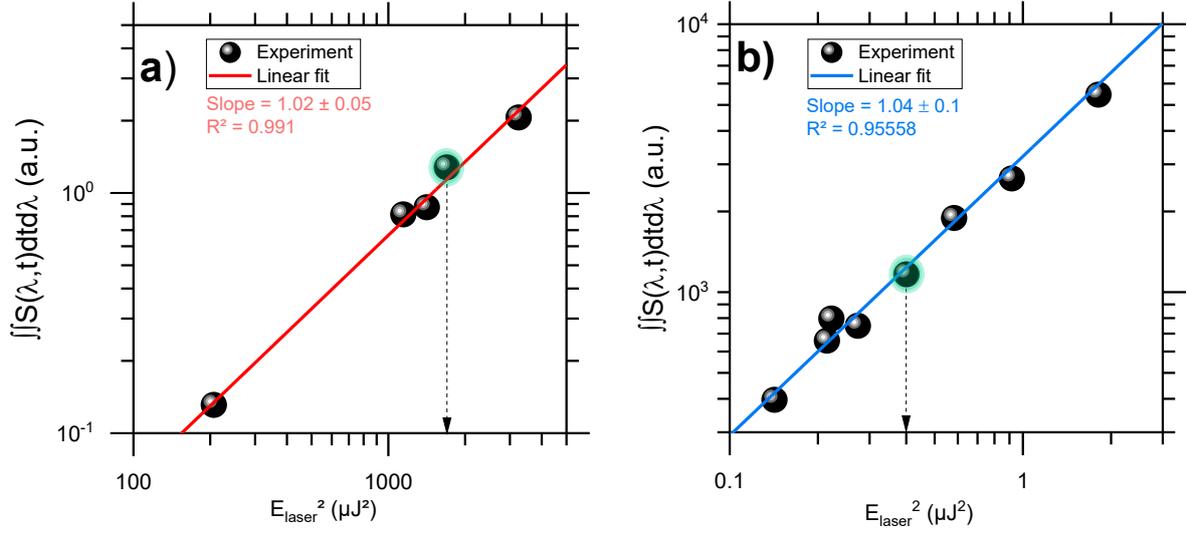

***Fig.10***: *Integrated (spectrally, temporally and spatially) TALIF versus the squared laser energy for a) H–atoms and b) Kr–atoms (P=200 Pa). The points highlighted green indicate the selected laser energies for performing a) H–atoms absolute density measurements and b) density calibration.*

The influence of $E_{Laser}$ on the temporally– and spectrally–integrated TALIF signals for H–atoms and Kr–atoms is shown in **Fig.10**. In the range of energies studied for each atom, a linear dependency (slope≈1) between the corresponding fluorescence intensity and $E_{Laser}^2$ is revealed, fulfilling the quadratic regime criterion (**eq.5**), in agreement with other relevant studies [**6–8,13–15,41**]. From these plots, the smallest possible laser energies in the quadratic regime are chosen to perform reliable TALIF measurements of H–atoms (plasma) as well as Kr–atoms (cuvette). The selected energies correspond to the data points highlighted green in **Fig.10** and are also given in **Tab.1**. Each point in the figure represents an integrated absorption line profile which is constructed by using temporally-integrated TALIF signals captured at different wavelengths around resonance with the streak camera. **Fig.11a1** and **Fig.11a2** show two indicative raw temporal TALIF signals recorded at resonance for the two energies highlighted in **Fig.10**. These are constructed through a spatial integration of raw signals within the same ROI of the CMOS sensor of the streak camera, as shown in **Fig.11a2** and **Fig.11b2**, respectively. Both images are rotated by 90° with respect to the default orientation of a streak image (cf. **Fig.8a**) for a more convenient representation. Besides, both signals are recorded using TR>>100 ps and exhibit slow rise times. The same slowing of the rising time can also be seen on the laser pulse itself (cyan) which appears distorted when captured at the same TR with TALIF. This, however, is not meaningful because the actual duration of the laser pulse is only 6 ps (**Fig.8**) and the two-photon absorption should take place within it. The origin of this distortion is the instrumental function of the streak camera. In fact, the larger the TR is, the slower the ramp applied to the sweep electrodes will be. This induces stronger repulsive forces between electrons of the photoelectron cloud in the streak tube and, thus, a more temporally-spread emission over the CMOS sensor's vertical axis [**9,10,42**]. Therefore, a methodology is developed to remove the instrumental function of the camera from these raw signals and obtain the actual TALIF. To this end, the instrumental function is determined using the laser pulse as a reference, the shape of which is distorted when increasing the TR beyond its minimum value, i.e., TR=100 ps (**Fig.8**). The distorted laser profile at different TR is indicative of the instrumental function of the streak. This is then deconvolved from the raw signals to obtain the actual TALIF and construct the absorption line profile (**Fig.12**). More details about this methodology can be found in [**9**]. The actual TALIF signals are also used for the determination of the effective lifetime of the atoms (**Fig.11a1** and **Fig.11b1**), which are needed for the absolute densities of H–atoms (**eq.5**).

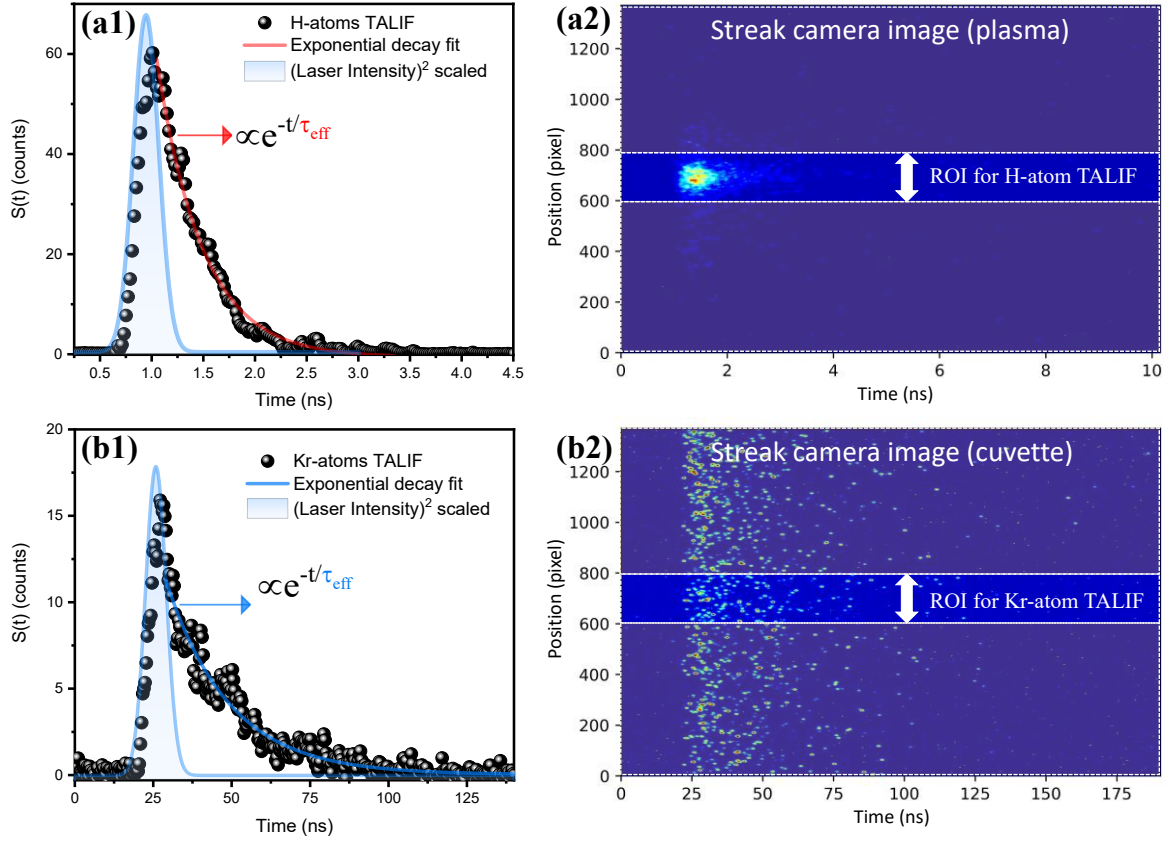

*Fig.11. Raw temporal ps-TALIF signals (black dots) of (a1) H–atoms (TR=5 ns, $\lambda_{Laser}$=205.06 nm, $E_{Laser}$=41.2 μJ, z=0 mm) and (b1) Kr–atoms (TR=200 ns, $\lambda_{Laser}$=204.11 nm, $E_{Laser}$=0.63 μJ), constructed from 11a2 and 11b2 (streak images), respectively. The corresponding distorted laser pulse profiles are shown in cyan (indicative of the instrumental function of the streak in each TR). After removing the instrumental function, the TALIF decays of the actual signals are used to determine the effective lifetime ($\tau_{eff}$) of the laser-excited states.*

The fluorescence spectral profiles of H–atoms and Kr–atoms corresponding to the energies selected in **Fig.10** are shown in **Fig.12a** and **Fig.12b**, respectively. Both profiles are approximated with Gaussian functions exhibiting FWHM of 29.5±10% pm and 27.5±10% pm, respectively. These linewidths agree with that provided by EKSPLA® (29 pm at 205 nm) being about 3 times larger than the theoretical value imposed by the transform limit (i.e., $\Delta\lambda_{Laser} \geq \frac{A\lambda_{Laser}^2}{c\Delta t_{Laser}}$ =10.3 pm; $\lambda_{Laser}$=205 nm, $\Delta t_{Laser}$ =6 ps, and A≈0.44 for a Gaussian beam [**43**]). The laser linewidth in our case is about 1.5 times larger than in ref. [**6**] (ps-TALIF studies of H–atoms in the COST plasma jet operating in He). Under their conditions, this was the dominant broadening mechanism of the absorption line, even though the profile was constructed by only 5 points due to limited wavelength resolution. Here we have twice the points for constructing the absorption line profile and the fitting still reveals a Gaussian profile in agreement with [**6**]. Since we also work at atmospheric pressure and our laser is spectrally broader, we expect that the line profile is strongly determined by the instrumental broadening of the laser, the other broadening mechanisms (Doppler and pressure) being significantly weaker. For instance, the Doppler broadening for H–atom is 1.4 pm ($\Delta\lambda_D = 7.16 \times 10^{-7} c/\lambda_0 \sqrt{T_G/M}$, c being the light speed, $\lambda_0 = \lambda_{Laser}/2$, $T_G$ the gas temperature and M the mass of H–atom). This agrees with the value of 1.3 pm found in [**6**] for $T_G$ =315 K, which is more than an order of magnitude smaller than that of the laser. Here the $T_G = 360 \pm 15$ K is estimated through the resonance broadening of the He I line at 667.8 nm using a line analysis method as in refs. [**44,45**].

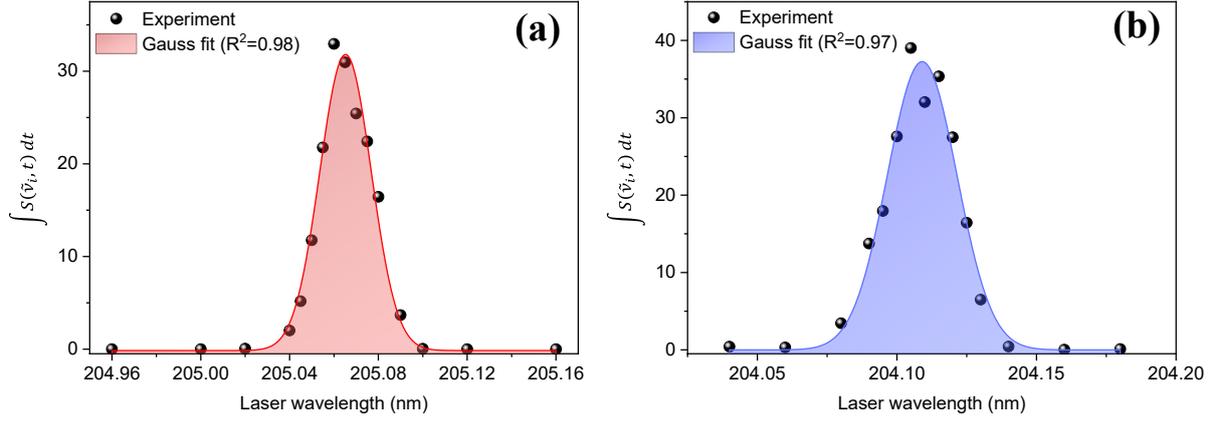

***Fig.12***: *Absorption line profiles (black dots) of (a) H–atoms and (b) Kr–atoms.*

Using the selected $E_{Laser}$ values along with the $t_{Laser}$ and $S_{Beam}$ measured, the laser intensities calculated (assuming a circular beam profile) are about 22 MWcm$^{-2}$ (Kr–atoms) and 1.4 GWcm$^{-2}$ (H–atoms). These are then considered to estimate the $W_{ge}$ and verify if the depletion of the atomic ground states by the laser is significant or not, since it could obscure the atomic density measurements. For H–atoms $W_{ge} = \frac{\sigma_H^{(2)} g(v_{res}) G^{(2)} I_{Laser}^2 \lambda_H^2}{(hc)^2}$ =3.37×10$^8$ s$^{-1}$, where $\sigma_H^{(2)}$ =1.77×10$^{-35}$ cm$^4$ [**2,32**], $G^{(2)}$ =2 [**2,3**], and $g(v_{res}) \simeq \frac{0.94}{FWHM} = \frac{0.94}{6.8\ cm^{-1}} \simeq$ 4.6×10$^{-12}$ s [**2**]. For Kr–atoms, the exact value of $\sigma_{Kr}^{(2)}$ is not available, the other values needed for $W_{ge}$ being measured in this study (as for H–atoms). Thus, we can only roughly approximate the $\sigma_{Kr}^{(2)}$ based on the ratio $\frac{\sigma_{Kr}^{(2)}}{\sigma_H^{(2)}}$ = 0.62 [**34**], which gives $W_{ge}$ = 5.1×10$^5$ s$^{-1}$. From the calculated $W_{ge}$ for each atom and the measured $t_{Laser}$ (**Fig.8**), the product $W_{ge} \times t_{Laser}$ equals to 3.1×10$^{-7}$ for Kr–atoms and 2×10$^{-3}$ for H–atoms. Thus, the condition $W_{ge} \times t_{Laser} \ll 1$ is fulfilled for both species.

Furthermore, the depletion rates of the laser excited states due to spontaneous relaxation (fluorescence) and quenching is qualitatively compared to that caused by PIN and SE. The effect of stimulated emission is not expected to be important under our conditions due to the narrow laser pulse and small interaction length of the laser's beam with the plasma/cuvette [**46**]. The negligible effect by SE is also suggested by a study performing ps-TALIF ($t_{Laser}$ =35 ps) and SE measurements in a H$_2$/O$_2$ flame, showing that SE only forms during a time interval which is comparable with the laser pulse duration [**47**]. For the effect of PIN, the depletion rates of the laser-excited states of both species can be estimated using the following formula [**47,48**]: $\Gamma_{PIN} = \frac{\sigma_{PIN} I_{Laser} \lambda_{Laser}}{hc}$, where $\sigma_{PIN}$ equals to 3.7×10$^{-23}$ m$^2$ and 3.6×10$^{-22}$ m$^2$ for the fluorescing states of H–atoms and Kr–atoms, respectively. This results to $\Gamma_{PIN}$ =5.3×10$^8$ s$^{-1}$ and $\Gamma_{PIN}$ =8.1×10$^7$ s$^{-1}$, respectively. However, for H–atoms the decay time of the fluorescing state can reach down to about 100 ps or lower at atmospheric pressure conditions [**7,47**]. This corresponds to a maximum decay rate of ~10$^{10}$ s$^{-1}$ which is significantly larger than the $\Gamma_{PIN}$ estimated here, thus denoting an insignificant depletion of the laser-excited state by PIN and SE compared to fluorescence and quenching. For laser-excited Kr 5p'[3/2]$_2$, the decay rate ($A$) is extracted from signals such as that shown in **Fig.11b1** recorded with the streak camera at different Kr pressures/densities in the cuvette. This is expressed as follows:

$$A = 1/\tau_{eff} = 1/\tau_{nat_{Kr}} + k_{Kr} \times N_{Kr} \quad (6)$$

with $k_{Kr}$ =1.46×10⁻¹⁰ cm³s⁻¹ [34] being the quenching rate of Kr 5p'[3/2]₂ and $N_{Kr}$ is the density of Kr–atoms (quenchers) in the cuvette.

The influence of Kr–atoms density on the decay rate of Kr 5p'[3/2]₂ state (**eq.6**) is shown in **Fig.13** (Stern–Volmer plot). For the pressures studied, the decay rate varies between 10¹⁰ s⁻¹ and 2×10¹⁰ s⁻¹, being again significantly larger than the corresponding $\Gamma_{PIN}$ estimated before. Furthermore, **Fig.13** allows determining the natural lifetime of excited Kr–atoms ($\tau_{nat_{Kr}} = 1/A_{nat}$) by performing a linear regression (orange line) on the experimental data and finding the intersection point with the vertical axis. The same figure also compares our results with different studies from the literature reporting $\tau_{nat_{Kr}}$ values ranging from 26.9±0.3 ns to 35.4±2.7 ns (see intersection points of the gray dashed lines with the vertical axis, respectively) [**34,49–51**]. The data corresponding to these works (shown in **Fig.13**) are calculated via **eq.6** using their natural lifetimes and the $k_{Kr}$ reported in ref. [**34**]. Our experiments here give $\tau_{nat_{Kr}}$ =31.9±2 ns which is close to the other reported values and is used for the calculation of $A_{Kr}$ in **Tab.1**.

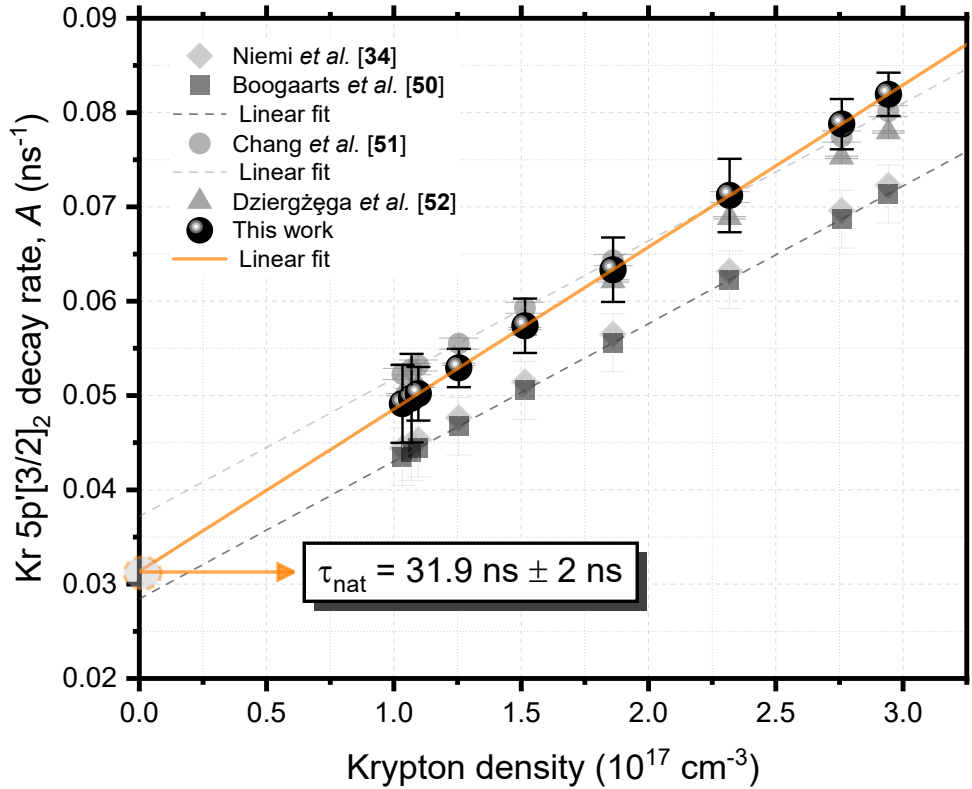

**Fig.13**: *Decay rates (A) of the Kr 5p'[3/2]₂ state density (black spheres) measured in this work against the Kr density in the cuvette. Error bars refer to the standard deviation from 5 independent experiments. These are compared with different studies from literature.*

### 3.3. *H–atoms effective lifetime and density*

The variation along the jet axis of the effective lifetime $\tau_{eff_{H(n=3)}}$ of the *H(n=3)* and of the absolute density of ground state H–atoms ($N_H$) is shown in **Fig.14**. Their spatial evolution is studied under two different gas flow rates: $Q_{He}$=0.3 and $Q_{He}$=0.1 slm. For these measurements, the electrodes are biased by the HV waveform shown in **Fig.3b1** while the instant of the laser pulse arrival relative to the onset of the HV pulse is set to $t_r$ (**Fig.3b1**). The choice of a constant instant $t_r$ is done because when changing the instant of the laser pulse's arrival relative to the trigger signal, the corresponding measured densities of H–atoms remain unchanged. This means that the atomic density has already reached a steady state, and no more temporal variations can be recorded. Therefore, with this configuration we can only study the influence of the gas flow rate and distance ($z$) along the jet axis on the steady state H–atoms density.

**Fig.14(a)** shows the dependence of the experimentally-measured $\tau_{eff_{H(n=3)}}$ on the gas flow rate and the axial distance $z$ from the tube's exit. For both flow rates, the measured lifetimes lie in the sub-ns timescale reaching values as low as 50 ps. These can still be experimentally measured with the streak camera due to its high temporal resolution. In both cases, the effective lifetime of *H(n=3)* is maximal near the tube's nozzle (370±50 ps for $Q_{He}$=1 slm and 320±25 ps for $Q_{He}$=0.3 slm) and declines with increasing distance, reaching down to 100±5 ps at $z$=4.5 mm and 50±10 ps at $z$=2 mm, respectively. The maximum distance for which a lifetime is still measurable at each flow rate correlates well with the corresponding jet's propagation length in **Fig.7** and **Fig.8**. Furthermore, in **Fig.8** the plasma effluent has a conical shape, i.e., it is thicker close to the exit nozzle and becomes narrower as we move upwards along the jet. This implies a weaker helium gas channel with increasing axial distance implying an increased penetration of the ambient air species ($N_2$, $O_2$, $H_2O$) into it, and a faster quenching of *H(n=3)* at larger distances. The decrease of $\tau_{eff_{H(n=3)}}$ with distance is more drastic for $Q_{He}$=0.3 slm compared to $Q_{He}$=1 slm. Indeed, by simply performing a linear regression on the experimental data shown in **Fig.14a**, $\tau_{eff_{H(n=3)}}$ falls down with a rate of ≈65 ps/mm at $Q_{He}$=1 slm which is about 3 times smaller than at $Q_{He}$=0.3 slm. Due to the 3–fold smaller gas flux at $Q_{He}$=0.3 slm compared to $Q_{He}$=1 slm, the corresponding plasma jet is expected to be weaker and more radially-spread, thus leading to a stronger quenching of the laser-excited atomic state by air species. Similar tendencies are obtained for the $N_H$ (**Fig.14b**) with maximum average densities measured close to the nozzle: $5.5\times10^{14}$ cm$^{-3}$ ($Q_{He}$=1 slm) and $4\times10^{14}$ cm$^{-3}$ ($Q_{He}$=0.3 slm). Furthermore, the maximal distances for which the effective lifetimes and atomic densities of H–atoms are still measurable are $z$=4.5 mm and $z$=2 mm for $Q_{He}$=1 slm and $Q_{He}$=0.3 slm, respectively. Both are very close to the corresponding propagation lengths (L) of the ionisation wave measured with the ICCD camera (**Fig.7**). The fact that $\tau_{eff_{H(n=3)}}$ and $N_H$ are not detected for larger distances in both cases implies that either the discharge is not capable of generating H–atoms or the density of H–atoms is below the detection limit of our system (around $10^{14}$ cm$^{-3}$). This threshold is mostly imposed by the dynamic range of the streak camera which is much smaller compared to those of conventional detectors usually used in TALIF experiments.

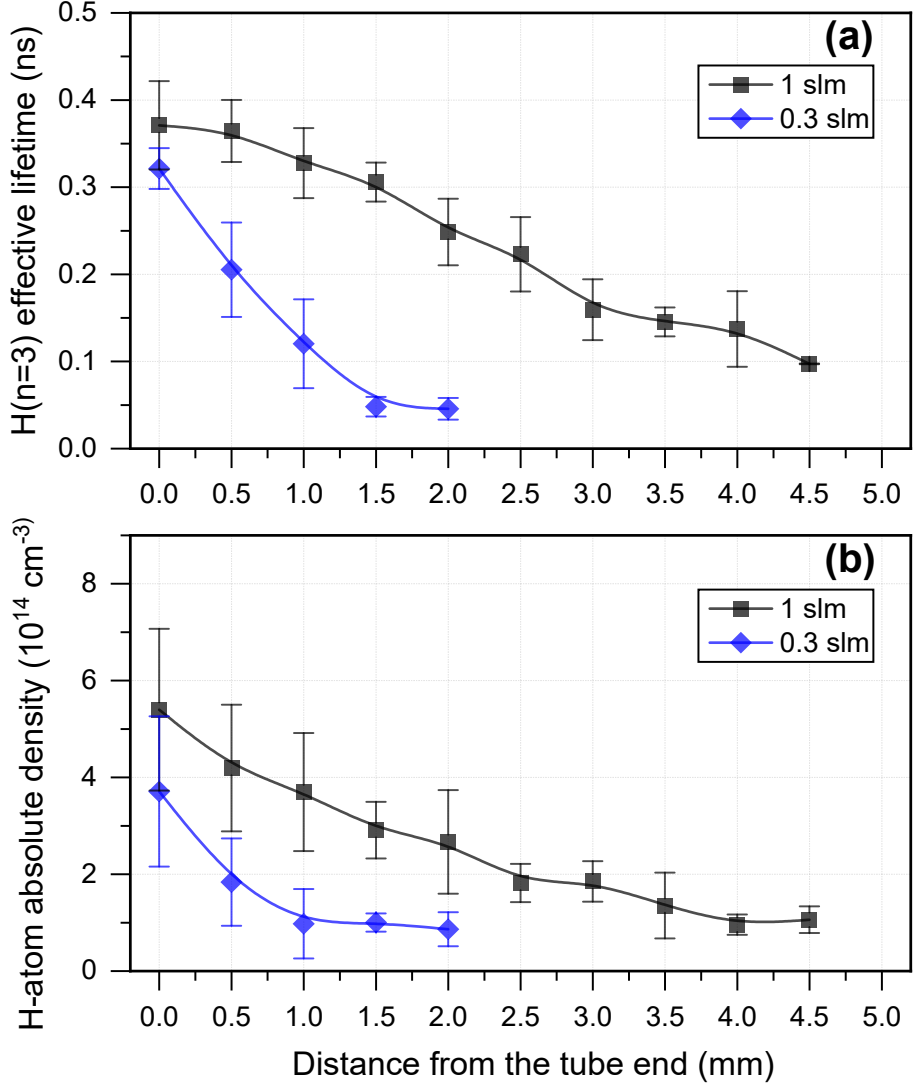

***Fig.14***. *Evolution of (a) the effective lifetime of the H(n=3) and (b) the absolute density of ground-state H–atoms along the jet's propagation axis for $Q_{He}$=0.3 slm and $Q_{He}$=1 slm. To determine the densities in (b), the average lifetimes from (a) at the corresponding distances are used. The error bars in (a) and (b) refer to the standard deviation (statistical error) obtained from 3 sets of experiments performed during different dates.*

To be able to investigate the temporal variation of $N_H$ in the time interval between two consecutive bursts applied to the μAPPJ, and also to follow the density built up during the very first HV pulses of a burst, the voltage waveform applied to the electrodes is modified (**Fig.3b2**). The modified HV waveform is also shown in **Fig.15** (grey) along with the measured densities corresponding to the two gas flow rates studied. Here it has to be mentioned that operating the power supply in burst mode did not allow it to ignite plasma for an electrode distance of 10 mm (**Fig.8**). Therefore, we were obliged to decrease the electrode distance (below 6 mm in this case) to sustain a stable discharge. This modification also resulted in an increase in the final density measured at z=0 mm for both gas flow rates compared to the corresponding data shown in **Fig.14**. However, this was the only way to study the temporal dynamics of the absolute density and the lifetime of ground state atomic hydrogen. From **Fig.15**, in the time interval between the two consecutive bursts shown (i.e., when the HV amplitude is zero), the density decreases almost linearly for both gas flow rates. Besides, a more drastic decrease is revealed for $Q_{He}$=1 slm with a decay rate of about $6\times10^{13}$ cm$^{-3}$/ms compared to $3\times10^{13}$ cm$^{-3}$/ms for $Q_{He}$=0.3 slm. Consequently, ground state H–atoms exhibit a larger residence time ($t_{res}$) in the helium gas channel for $Q_{He}$=0.3 slm ($t_{res}$≈1.2 ms) compared to that measured for $Q_{He}$=1 slm ($t_{res}$≈0.5 ms). Similarly, during the initial HV pulses of the following burst, a larger growth rate of the density is measured for $Q_{He}$=1 slm (about $3.5 \times 10^{13}$ cm$^{-3}$/ms) than for $Q_{He}$=0.3 slm (about $2.5 \times 10^{13}$ cm$^{-3}$/ms), and the corresponding density reaches relatively faster (within about 15 HV periods) at steady state conditions than for $Q_{He}$=0.3 slm (within about 20 HV periods).

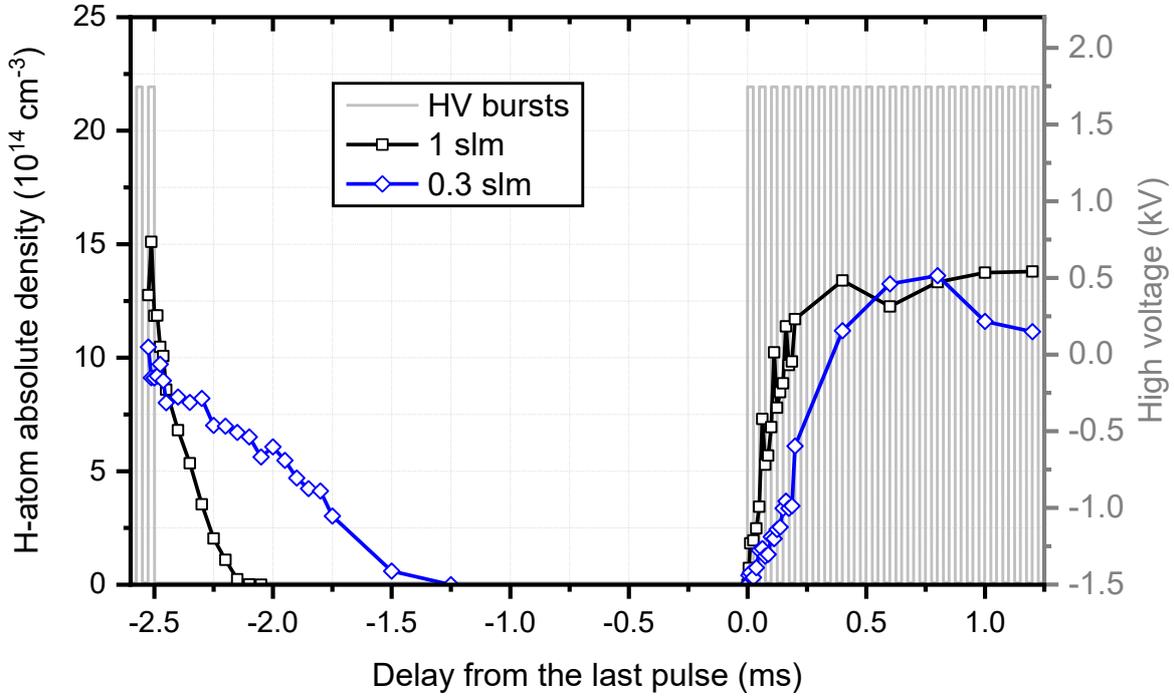

**Fig.15**. *Temporal variation of the absolute density of H–atoms at z=0 mm between two bursts (i.e., plasma "OFF" time) and during various pulses within a burst for $Q_{He}$=0.3 slm and $Q_{He}$=1 slm.*

## 4. Discussion

This work deals with a μm scale (≈450 μm tube radius) helium plasma jet penetrating the room air which is mainly composed of $N_2$ (≈78%) and $O_2$ (≈21%). H–atoms in our conditions can originate in the water vapour also present in the room (50% relative humidity at 20 °C) and the impurities contained in the gas bottle (≤3 ppm $H_2O$, ≤0.5 ppm $C_nH_m$). A mechanism that could be responsible for the H–atoms generation is the dissociative excitation of water by direct electron impact [13,52,53]:

$$H_2O(X) + e^- \rightarrow OH(A) + H + e^- \quad (7)$$

As a matter of fact, the formation of OH(A) is evidenced by the presence of OH(A–X) emission in **Fig.8**. The electron density in our system can be empirically estimated from the discharge's current density (*J*), as follows:

$$n_e = \frac{J}{ev_d} = \frac{I_p}{ev_d \pi R^2} \quad (8)$$

where *e* is the elementary charge, $v_d$ is the electrons' drift velocity assumed to be similar to the average velocity of the ionisation waves (**Fig.7**), and *R* is the tube radius. Using the discharge's peak current ($I_P$≈3.5 mA) recorded at the conditions of **Fig.6** and for an average $v_d$≈0.025 mm/ns (**Fig.7**; $Q_{He}$=1 slm) we roughly obtain $n_e \approx 5\times10^{13}$ cm$^{-3}$. Other processes involving electrons may refer to dissociative recombination of $H_2O^+$ and destruction of protonated water clusters [6].

Furthermore, atomic hydrogen can be formed through water dissociation by VUV radiation (also producing OH), water dissociation by He metastables, and OH dissociation by atomic oxygen [6,13,40,54]. Klute et al. showed that a similar µAPPJ device can efficiently generate VUV radiation in the range 55–180 nm originating in emissions from 1st and 2nd $He_2^*$ excimer and other species [24]. Although this makes their contribution possible, it is not feasible to perform a quantitative map of their distribution in the effluent and better assess their role. Furthermore, the influence of He metastables is not expected to be significant for densities on the order of $10^{11}$ cm$^{-3}$ as was the case for the COST-Jet [54]. However, in our system their density is expected to be on the order of $10^{13}$ cm$^{-3}$ [55], their effective lifetime reaching several microseconds in atmospheric pressure He plasma jets [55,56]. This increases the probability of their involvement in the H–atom production, especially in the time interval after the rising and falling phases of the applied voltage. Finally, excited O–atoms are identified in the emission spectrum (**Fig.8)**, indicating the presence of ground state O–atoms which can interact with OH and dissociate it to create H–atoms. The relative contribution of these reactions also depends on the water content in the jet [6]. At 20 ºC (this work), the saturation vapour pressure of water is 23.4 mbar which corresponds to an $H_2O$ density of $5.7\times10^{17}$ cm$^{-3}$ at atmospheric pressure conditions. For a relative humidity of 50% (this work), the water vapour pressure is 11.7 mbar which corresponds to a density of about $3\times10^{17}$ cm$^{-3}$. However, it is difficult to have an accurate estimate of the water quantity which is dragged into the jet for the two gas flow rates studied.

For the loss mechanisms of ground state H–atoms, diffusion can be a possible mechanism. In a kHz modulated RF-driven plasma jet generated in a gas mixture of He with 0.1%$H_2$ (2 slm total flow rate; 1 mm inner tube radius), radial diffusion was among the responsible mechanisms for the decrease in H–atoms. This was due to a non-negligible diffusion time ($t_{diff}$=1 ms) against the gas residence time in the tube (0.5 ms) [57]. In **Fig.15** the experimentally-measured $t_{res}$ of the ground state H–atoms are about 0.5 ms and 1.2 ms for $Q_{He}$=1 slm and $Q_{He}$=0.3 slm, respectively, being close to the values reported in [57]. The effective diffusion coefficient of H–atoms ($D_{eff}$) can be roughly estimated as:

$$D_{eff} = \frac{L^2}{t_{diff}} \quad (9)$$

where $L$ is an effective mixing length, here being considered to be equal to the tube diameter (450 µm). If we simply consider that $D_{eff} = D_{H \to He}$, we obtain $t_{diff} \approx 0.8$ ms ($D_{H \to He} \approx 2.42\times10^{-4}$ m$^2$/s being the diffusion coefficient of H–atoms in helium for a gas temperature $T_G$=333 K [58], which is close to the estimated $T_G$=360±15 K in our work). The calculated $t_{diff}$ is relatively close to the lifetimes obtained in **Fig.15**. The small differences could be due to the contribution from other loss channels of H–atoms as well as the different mixing of air species into the helium channel between the two flow rates. Additional consumption mechanisms of H–atoms could be through collisions with neutral species in the plasma (such as He, $O_2$ and $O_3$) [6,57], Penning ionisation with He* and $He_2^*$, electron impact ionisation and charge exchange with $O^+$ [13]. The relative impact of these channels requires a dedicated numerical simulation and cannot be accessed in this work.

Furthermore, the experimentally-measured values of $\tau_{eff_{H(n=3)}}$ for both flow rates drop with increasing distance from the exit nozzle (**Fig.14a**). This is attributed to a strong mixing of the helium jet with the ambient air. The reduction in $\tau_{eff_{H(n=3)}}$ is more drastic for $Q_{He}$=0.3 slm compared to $Q_{He}$=1 slm. The quenching of H(n=3) is essentially driven by collisions with He, $N_2$, $O_2$, and $H_2O$. The relative significance of each reaction also depends on the density of the corresponding collider in the helium jet. Different sets of quenching coefficients $k_X$ ($X$: $N_2$, $O_2$, $H_2O$ or He) of H(n=3) have been published in the literature as shown in **Tab.2**. The last column of **Tab.2** shows the weighted mean values calculated for each quencher based on a combination of the data given in the relevant references. The weighted mean coefficients can then be utilised to calculate a theoretical $\tau_{eff_{H(n=3)}}$. This is done by considering different relative densities of air species and He atoms in the jet, starting with the highest total density of about $2.5 \times 10^{19}$ cm$^{-3}$ (20 °C and atmospheric pressure). For this *i)* we define the total air density as $N_{air}=N_{N_2}+N_{O_2}+N_{H_2O}$ with $N_{N_2}=0.78 \times N_{air}$, $N_{O_2}=0.21 \times N_{air}$ and $N_{H_2O} \approx 10^{-4} \times N_{air}$ (assuming that 1% of the water vapour pressure is dragged into the jet), and *ii)* we simply vary the total percentage of air (%$N_{air}$) in the He gas from 0% to 50%.

***Tab.2***: *Published quenching coefficients of H(n=3) by He, $N_2$, $O_2$, and $H_2O$.*

| Quencher $X$ | $k_X$ in $10^{-10}$ cm$^3$/s ($\pm$ error) | | | | |
|---|---|---|---|---|---|
| | Ref. | [36–38] | [33] | [34] | [38] | Weighted mean values |
| He | 0.099 (0.05) 0.317 (0.002) | 0.53 (0.005) | 0.18 (-) | – | 0.2692 (0.0014) |
| $N_2$ | 27.7 (1.1) | 25.4 (17.3) | 20.1 (-) | – | 23.9030 (0.777) |
| $O_2$ | 26 (1) | 36.5 (4.2) | 32.6 (-) | – | 29.4985 (0.6973) |
| $H_2O$ | 110 (10) | – | – | 91 (16) | 104.6629 (8.48) |

Then, the theoretical $\tau_{eff_{H(n=3)}}$ is calculated as follows:

$$\frac{1}{\tau_{eff_{H(n=3)}}} = \frac{1}{\tau_{nat_{H(n=3)}}} + \sum_X k_X N_X \quad (10)$$

where $\tau_{nat_{H(n=3)}}$ is obtained as explained in **section 2.3** and $N_X$ is the hypothetical relative density of different air species penetrating the He jet. This calculation allows directly comparing the theoretical $\tau_{eff_{H(n=3)}}$ with the experimentally-measured values shown in **Fig.14a**. It also allows estimating the corresponding amounts of air penetrating the jet at each axial position and flow rate as shown in **Fig.16**.

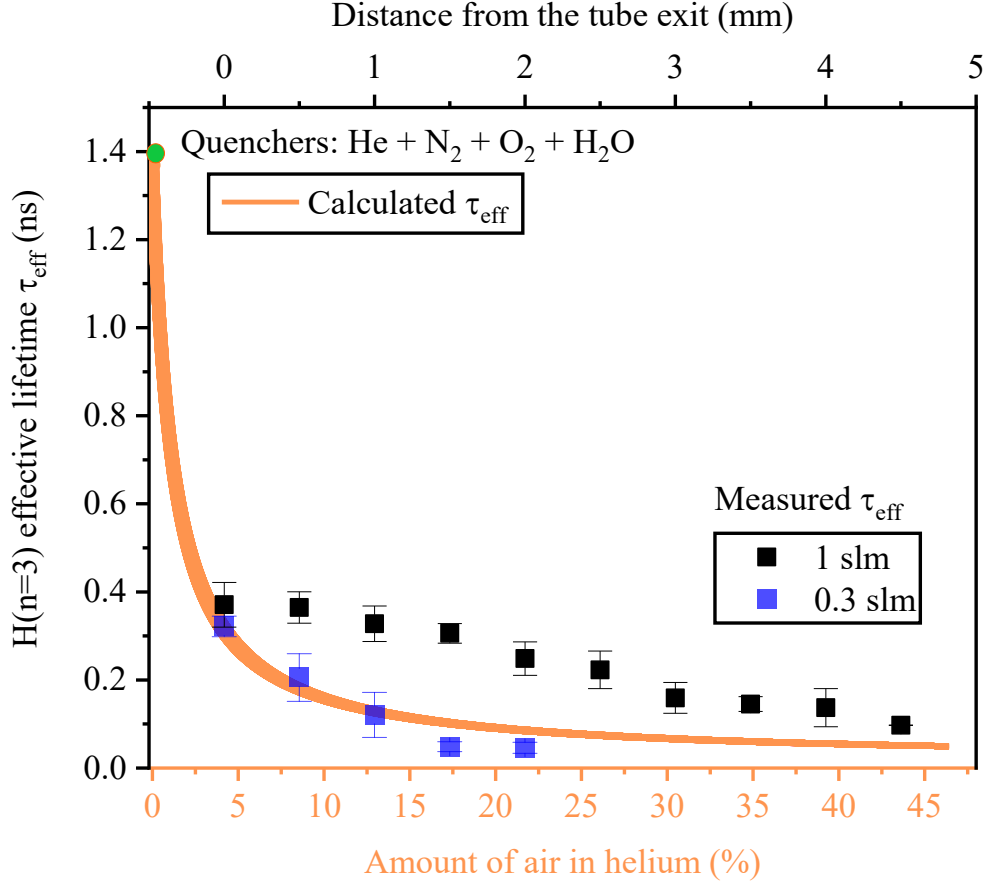

***Fig.16***. *Weighted mean $\tau_{eff_{H(n=3)}}$ (eq.10; orange) versus the air percentage in the jet (bottom axis). The experimentally-measured $\tau_{eff_{H(n=3)}}$ along the jet axis (top) is also shown (taken from **Fig.14a**).*

**Fig.16** shows the dependence of the calculated theoretical $\tau_{eff_{H(n=3)}}$ (*eq.10*; orange line) on the air concentration into the plasma jet. The black and blue squares are representative of the variation of the experimentally-measured $\tau_{eff_{H(n=3)}}$ along the jet axis for the two flow rates studied. The projection of the experimental lifetimes on the orange line and of the corresponding calculated values on the horizontal (bottom) axis provide an estimate of the air percentage into the jet which induces the quenching of H(n=3). Here, the He atoms alone cannot induce a substantial quenching of H(n=3). This is because when considering only He atoms as quenchers (i.e., considering a maximum $N_{He}$=2.5×10$^{19}$ cm$^{-3}$), e.g., at the exit nozzle (z=0 mm), the calculated theoretical $\tau_{eff_{H(n=3)}}$ is about 1.4 ns (green dot in **Fig.16**). This value is much larger than the maximum experimentally-measured $\tau_{eff_{H(n=3)}}$ for each flow rate (370±50 ps for $Q_{He}$=1 slm and 320±25 ps for $Q_{He}$=0.3 slm). At the same axial position (z=0 mm), the calculated theoretical $\tau_{eff_{H(n=3)}}$ for both flow rates match with the experimentally-determined values only when considering an air concentration into the jet of about 3%. A rather similar mixing of the helium channel with the air species is obtained at z=0 mm for the two flow rates. However, for z>0 mm the air penetration into the jet quickly becomes critical for $Q_{He}$=0.3 slm, reaching >40% at z=2 mm and causing a significant quenching of H(n=3) ($\tau_{eff_{H(n=3)}}$=50±10 ps). Contrary, for $Q_{He}$=1 slm the air percentage into the jet at z=2 mm remains much smaller (about 6%) resulting in 5 times larger $\tau_{eff_{H(n=3)}}$. The smaller air entrainment in the He channel at higher axial distances for $Q_{He}$=1 slm compared to $Q_{He}$=0.3 slm is also evident from the larger propagation length of the corresponding jets (**Fig.7** and **Fig.8**).

## 5. *Conclusions*

This study was devoted to the accurate experimental determination of the effective lifetime ($\tau_{eff_{H(n=3)}}$) and absolute density ($N_H$) of atomic hydrogen in a helium atmospheric pressure microplasma jet (μAPPJ). Generated ground-state H–atoms were excited to H(n=3) using picosecond two-photon laser induced fluorescence (ps–TALIF). Absolute density values were calibrated by performing identical ps–TALIF measurements in krypton gas contained in a custom-built low-pressure cuvette. To capture the rapid decay times of the fluorescence signals, a streak camera (~1 ps highest temporal resolution) was employed. Its use also allowed for the direct measurement of the laser pulse duration (≈6 ps around 205 nm), a key parameter defining the time resolution of the measurements of fluorescence decay times as well as the laser intensity which affects the fluorescence yield.

The successful combination of ps–TALIF with the streak camera allowed us to directly resolve sub-nanosecond values of $\tau_{eff_{H(n=3)}}$ in the effluent of the μAPPJ. The measured $\tau_{eff_{H(n=3)}}$ increased with the helium flow rate ($Q_{He}$) in the range between 0.3 and 1 slm, being smaller than 400 ps near the exit nozzle of the μAPPJ and dropping below 100 ps at axial distances as short as 2 mm ($Q_{He}$=0.3 slm) and 5 mm ($Q_{He}$=1 slm). These variations highlighted a critical role of ambient air entrainment and collisional quenching in shaping the spatial evolution of fluorescence signals. The measured effective lifetimes were essential for the determination of $N_H$, which also increased with the gas flow rate. $N_H$ was maximised close to the exit nozzle (5.5×10$^{14}$ cm$^{-3}$ and 4×10$^{14}$ cm$^{-3}$ for $Q_{He}$=1 and 0.3 slm, respectively) and minimised (around 10$^{14}$ cm$^{-3}$) at 2 mm ($Q_{He}$=0.3 slm) and 5 mm ($Q_{He}$=1 slm). A large systematic uncertainty on the measured density as high as 64% (depending on the operating condition) was inevitable. This was mainly induced by the ratio of the two-photon absorption cross sections of Kr and H atoms ($\sigma^{(2)}{}_{Kr}/\sigma^{(2)}{}_H$) which has not been measured yet in the picosecond regime. Shortening the electrode distance resulted in an increased $N_H$ being as high as 1.5×10$^{15}$ cm$^{-3}$ near the exit nozzle. In this configuration, the μAPPJ was operated in burst mode which allowed estimating the residence time ($t_{res}$) of ground state H–atoms in the helium gas channel. This was found to be $t_{res}$≈1.2 ms for $Q_{He}$=0.3 slm and $t_{res}$≈0.5 ms for $Q_{He}$=1 slm. Ground-state H–atoms can be consumed via different mechanisms such as diffusion, collisions with neutral ground-state species, and metastables among others.

This system enables accurate measurements of atomic species' effective lifetimes under conditions where conventional nanosecond diagnostics (ns–TALIF, ICCD) reach their limits. Particularly, it provides important insights into the spatio-temporal dynamics of atomic hydrogen in microplasma jets which can be essential for applications in surface processing, analytical chemistry, and/or biomedicine. Combining these ultrafast diagnostics with numerical models and/or additional diagnostics (e.g., absorption spectroscopy to probe helium metastables) could further illuminate the interplay between plasma dynamics, gas-phase chemistry, and radical transport in atmospheric-pressure environments.

## *Acknowledgments*

This work was funded by the ANR ULTRAMAP Project (ANR–22–CE51–0027), the Labex SEAM Project (ANR–10–LABX–0096; ANR–18–IDEX–0001), and the IDF regional Project SESAME DIAGPLAS. The authors would like to acknowledge (i) Mr Ludovic William for his contribution to the fabrication of the custom-built cuvette, and (ii) Mr Odhisea Gazeli for his assistance in the preparation of Fig.1.

*APPENDIX*

*H–atom absolute densities and related systematic errors*

An error analysis is conducted to evaluate the systematic error associated with the measured densities of H–atoms ($N_H$), following the standard error propagation approach. In this work, the systematic error in the $N_H$ directly depends on the systematic uncertainties of the different quantities used in **eq.5**. For TALIF experiments, the variations of these quantities are small and independent and, thus, the differentials can be approximated by finite variations by neglecting higher-order terms. In this context, considering that $N_H$ is a function of $i$ independent variables $x_i$, each with an uncertainty $\sigma(x_i)$, the total uncertainty in the density, $\sigma(N_H)$, is given by:

$$\sigma(N_H) = \sqrt{\sum_i \left(\frac{\delta_{N_H}}{\delta_{x_i}}\sigma(x_i)\right)^2}$$

This approximation simplifies the determination of the standard deviation, which is obtained by taking the square root of the sum of the squared contributions from all quantities included in **eq.5**. For functions involving products or quotients, this can be rewritten in terms of relative uncertainties:

$$\sigma(N_H) = \sqrt{\sum_i \left(a_i \frac{N_H}{x_i}\sigma(x_i)\right)^2} = N_H \sqrt{\sum_i \left(a_i \frac{\sigma(x_i)}{x_i}\right)^2}$$

where $a_i$ denotes the exponent of a quantity of **eq.5**. This leads to a practical approach: rather than working with partial derivatives, one can directly take the square root of the sum of the squared relative uncertainties and multiply it by $N_H/100$ to obtain the absolute systematic uncertainty as follows:

$$\sigma(N_H) = \frac{N_H}{100}\sqrt{\sum_i (a_i \triangle_i)^2}$$

with

$$\triangle_i = 100 \frac{\sigma(x_i)}{x_i}$$

represents the relative uncertainty of $x_i$ expressed as a percentage. Applying this to the quantities of **eq.5**, the absolute systematic error $\sigma(N_H)$ is determined as follows:

$$\sigma(N_H) = \frac{N_H}{100}\sqrt{\triangle^2_{N_{Kr}} + \triangle^2_{S_{F_{Kr}}} + \triangle^2_{S_{F_H}} + \triangle^2_{\tau_{effKr}} + \triangle^2_{\tau_{effH}} + \ldots}$$

$$\sqrt{\ldots + \triangle^2_{g(v_{res})_{Kr}} + \triangle^2_{g(v_{res})_H} + \triangle^2_{A_{Kr}} + \triangle^2_{A_H} + \ldots}$$

$$\sqrt{\ldots + \triangle^2_{\eta_{Kr}} + \triangle^2_{\eta_H} + \triangle^2_{T_{Kr}} + \triangle^2_{T_H} + \triangle^2_{\sigma^{(2)}_{Kr}/\sigma^{(2)}_H} + (2\triangle_{E_{Kr}})^2 + (2\triangle_{E_H})^2}$$

This method is particularly useful when dealing with experimental measurements and complex equations where the relative uncertainties are more readily available than the absolute ones. The main systematic errors associated with the quantities of **eq.5** are provided in **Tab.1**. As an example of application of the method, the systematic error $\sigma(N_H)$ corresponding to the measured $N_H$ at $z=0$ mm and $Q_{He}=1$ slm is 64%.

The factor inducing the largest density error is the ratio $\sigma^{(2)}_{Kr}/\sigma^{(2)}_{H}=0.62\pm0.31$ [34] which has been measured using a narrowband nanosecond (ns) dye laser (10 ns pulse duration). This agrees with the value $\sigma^{(2)}_{Kr}/\sigma^{(2)}_{H}=0.56\pm28$ [49] obtained with a similar ns laser system (0.2 cm$^{-1}$ bandwidth around 205 nm). Recently, a new value of the ratio $\sigma^{(2)}_{Kr}/\sigma^{(2)}_{H}$ has been determined (0.027±20%) using a broadband femtosecond (fs) laser system (90 fs pulse duration, 200 cm$^{-1}$ bandwidth) [59]. This value is about 20 times smaller than the two previous ones obtained using ns laser systems. This important difference in the $\sigma^{(2)}_{Kr}/\sigma^{(2)}_{H}$ between ns and fs lasers could be attributed to the very different spectral widths of the laser lines used [59]. Concerning ps–TALIF, there is no published value yet for the corresponding ratio $\sigma^{(2)}_{Kr}/\sigma^{(2)}_{H}$. Based on the fact that the ps laser features (e.g., ≈6 ps pulse width and 6.8 cm$^{-1}$ bandwidth at 205 nm in this work) lie between those of ns and fs lasers, one would expect an intermediate $\sigma^{(2)}_{Kr}/\sigma^{(2)}_{H}$ value which has yet to be verified. An accurate and independent determination of the ratio $\sigma^{(2)}_{Kr}/\sigma^{(2)}_{H}$ in each regime (i.e., ns–, ps– and fs–TALIF) would be ideal and could significantly improve the accuracy of the corresponding TALIF density measurements. Nevertheless, the value $\sigma^{(2)}_{Kr}/\sigma^{(2)}_{H}=0.62\pm0.31$ [34] was used in this work since it has been adopted in several previous and recent studies to determine absolute H–atom densities in plasmas using ps– and fs–TALIF [6,7,13,15,36].